\documentclass[preprint,longbibliography,aps,pra,superscriptaddress,showpacs,floatfix]{revtex4-1}
\usepackage{graphics,epsfig,graphicx}
\usepackage{amsbsy}
\usepackage{bm}
\usepackage{amsfonts}
\usepackage{cancel}
\usepackage{multirow}
\usepackage{amsmath}
\usepackage{wasysym}
\usepackage{xcolor}
\begin{document}

\title{Subleading contributions to $N$-boson systems inside the universal window}

\author{P. Recchia}
\affiliation{
 Universit\'e C\^ote d'Azur, CNRS, Institut  de  Physique  de  Nice,
1361 route des Lucioles, 06560 Valbonne, France }
\author{A. Kievsky} 
\affiliation{Istituto Nazionale di Fisica Nucleare, Largo Pontecorvo 3, 56100 Pisa, Italy}
\author{L. Girlanda}
\affiliation{Dipartimento di Matematica e Fisica "E. De Giorgi", Universit\`a del
Salento, I-73100 Lecce, Italy}
\author{M. Gattobigio}
\affiliation{
 Universit\'e C\^ote d'Azur, CNRS, Institut  de  Physique  de  Nice,
1361 route des Lucioles, 06560 Valbonne, France }
\begin{abstract}
	We study bosonic systems in the regime in which the two-body
	system has a shallow bound state or, equivalently, a large value of the two-body
	scattering length. Using the effective field theory framework
	as a guidance, we construct a series of potential
  terms which have a decreasing importance in the description of the
	binding energy of the systems. The leading order potential terms consist in a
  two-body term, usually attractive, plus a three-body term, usually repulsive;
  this last term is required to prevent the collapse of systems with more than
  two particles. At this order, the parametrization of the two-body potential is done to
  obtain a correct description of the scattering length, which governs the
  dynamics in this regime, whereas the three-body term fixes a three-body datum. 
  We investigate the role of the cut-off in the leading order
  description and we extend  the exploration beyond the leading order by
  including the next-to-leading order terms in both, the two- and three-body
  potentials. We use the requirement of the stability of the $N$-body system,
  whose energy is variationally estimated, to introduce the three-body forces.
  The potential parametrization, as a function of the cut-off, is fixed to
  describe the energy of $^4$He clusters up to seven particles within the 
  expected accuracy. Finally, we also explore the possibility to 
  describe at the same time the atom-dimer scattering length.  
\end{abstract}
\maketitle

\section{Introduction}
In the last years many efforts have been directed to the study of systems
existing at, or close to, the unitary limit, a limit in which the two-body scattering
length diverges. The interest is based on the universal
properties exhibited by such
systems~\cite{braaten:2006_PhysicsReports,naidon:2017_Rep.Prog.Phys.,%
kievsky:2021_Annu.Rev.Nucl.Part.Sci.}. The energy spectrum shows a
scale invariance and can be effectively described by a
limited number of parameters, typically the scattering length $a$ of the
two-body system and the three-body parameter $\kappa_*$, which gives the binding
energy, $E_*=\hbar^2 \kappa_*^2/m$, of the three-body system at the unitary point.

Examples of such systems come from nuclear physics, where the singlet- and
triplet-scattering lengths are both much greater than the typical interaction
length, and from atomic $^4$He, where the scattering length is much greater than the
van der Waals length $\ell_{\text{vdW}}$, which represents the typical
interaction length in atomic physics. 

These systems have been historically described by potential models. In nuclear
physics, the first phenomenological potentials have been 
derived by parametrizing the most general interaction allowed by the symmetries,
and explicitly incorporating the long-range part, i.e. the one-pion
exchange. The strength of the different potential terms
were fixed by a fitting procedure designed to reproduce as best as possible 
the two-body scattering data and the deuteron binding energy. In
atomic physics, and for helium in particular, the potential curves have been
constructed by a mix of {\it ab-initio} calculations, taking into account
the repulsive interaction between the electronic clouds, and empirical
parametrizations trying to incorporate as many experimental data as possible, as
the virials and the viscosity~\cite{aziz:1979_J.Chem.Phys.,aziz:1987_MolecularPhysics,aziz:1991_J.Chem.Phys.}. 
Also in these cases the van der Waals long distance behavior induced by the electric multipoles,
$\sum_\lambda C_\lambda/r^\lambda, \lambda=6,8,10$, is explicitly included.

Already in the old times, it had been realized that some low-energy observables
were insensitive to the details of the potentials. For instance, in nuclear
physics the $s$-wave two-body phase-shift up to energies of 10--15~MeV could be
reproduced by any two-parameter potential compatible with Bethe's effective
range expansion (ERE)~\cite{fermi:1936_Ricercasci.,bethe:1949_Phys.Rev.}. 
Other evidences are the correlation between the triton binding energy and the
neutron-deuteron doublet scattering length, known as Phillips
line~\cite{phillips:1968_Nucl.Phys.A}, and the correlation between the triton
and the alpha-particle binding energies, known as Tjon line~\cite{tjon:1975_Phys.Lett.B}. 
Correlations of this type have been explained by V. Efimov showing that
the three-nucleon system at low energies is governed by the three-body
parameter $\kappa_*$. Moreover, using a zero-range model V. Efimov predicted the
Efimov effect, a remarkable property of the three-body system 
located at the unitary
limit~\cite{efimov:1970_Phys.Lett.B,efimov:1971_Sov.J.Nucl.Phys.,efimov:1988_Few-BodySystems}.

The modern approach to describe these systems is the effective field theory
(EFT) framework, that exploits the small expansion parameter $\ell/a \ll 1$
describing  the  scale separation between the range of the interaction $\ell$
and the scattering length $a$. In nuclear physics, such a theory is known as
pionless
EFT~\cite{vankolck:1998_ChiralDynamics:TheoryandExperiment,vankolck:1999_Nucl.Phys.A,%
bedaque:1998_Phys.Rev.C,kaplan:1998_Phys.Lett.B,kaplan:1998_Nucl.Phys.B,birse:1999_Phys.Lett.B},
indicating that the pion degrees of freedom have been integrated out.  In this
case, the short-distance scale of the theory is the inverse of the pion mass
$\ell= 1/M_\pi\approx 1.4\,$fm. The same theory has been used to describe atomic
$^4$He~\cite{bazak:2016_Phys.Rev.A,bazak:2019_Phys.Rev.Lett.}, where the
short-distance scale is $\ell_{\text{vdW}}=5.08\,a_0$, with $a_0$ the Bohr
radius.  In order to use the EFT to compute observables, the theory must be
first regularized, for instance with a momentum cut-off $\Lambda$. The
renormalization procedure allows to reduce the dependence of physical
observables on $\Lambda$ at a level compatible with the neglected orders of the
small-parameter expansion. For so doing, a power counting has to be established
that allows to identify the operators entering at each order.  To ensure
approximate cut-off independence, the subleading interactions are to be treated
perturbatively, on the top of a non-perturbative treatment of the leading order (LO) 
interaction, mandated by the description of shallow bound states.

At the LO in the small parameter expansion of this EFT there are a two-body and 
a three-body force whose strengths, the low energy constants (LECs), can be
fixed to reproduce a two-body datum, usually the scattering length, and a
three-body datum, usually the ground-state trimer energy.  The promotion of the
three-body force to LO is a characteristic of this particular EFT and originates
from the necessity of stabilizing the systems with more than two particles
against the Thomas collapse~\cite{thomas:1935_Phys.Rev.}. An interesting
phenomenon that characterizes this kind of systems is the emergence of a
discrete scale invariance observed in three- and four-boson systems close to the
unitary limit, which reflects the existence of limit cycle in the
renormalization group flow.

The EFT can be used to inspire and organize the construction of potential models
representing the interaction between particles. In the present case of systems
with a large value of the two-body scattering length, these potentials capture
the system universal properties.  The potentials appear as a sum of terms ordered
according to the power counting of the EFT. Since the whole truncated potential
is treated non-perturbatively, the $\Lambda \rightarrow +\infty$ limit cannot be
taken; rather, the cutoff is to be mantained of the order of the short-distance
scale. Within this limited range, the dependence of observables on the cutoff
can still profitably be scrutinized.  Eventually, an optimized cutoff can be
chosen, to improve the description of a number of experimental data.

In particular, the following exploration can be carried out: after fixing the LO
potential by the two-body scattering length and by the binding energy of the
three-body system, the same potential can be used to compute the energy levels
for systems with increasing number of particles. We expect that the LO
description, accurate at the percentage level given by the ratio $\ell/a$,
remains inside the same percentage level as the number of particles increases.
An analysis of this kind has been started in
Refs.~\cite{kievsky:2017_Phys.Rev.A,kievsky:2018_Phys.Rev.Lett.,kievsky:2020_Phys.Rev.A},
where it has been shown that there is a small range of cut-off values that 
extends the validity of the LO description to larger systems. Here we further
analyze this fact and extend the study to consider the next-to-leading order
(NLO).  This term has been considered perturbatively in
Ref.~\cite{bazak:2019_Phys.Rev.Lett.} up to $N=6$ with the conclusion that a
subleading four-body force is needed to stabilize the systems with
$N>4$. 

In the present study we analyze the effects of the NLO potential terms in
the description of the $N=4,5,6,7$ systems and estimate the limit
$N\rightarrow\infty$. To this aim we consider a system of equal bosons
inside the universal window with the use of a Gaussian regulator at LO and NLO. 
The atomic $^4$He
system will be taken as a reference system to judge the quality of the
effective description. Very seldom experimental data exist for this
system with arbitrary number of particles. Essentially the dimer and
trimer binding energies were recently measured~\cite{kunitski:2015_Science}. 
So, we use reference data results obtained by one of the widely used helium-helium interaction, 
the LM2M2 potential. For the purpose of the present analysis the numerical
results of the LM2M2 for the binding energy of different clusters are
considered equivalent to experimental data. With the inspired EFT
potential, we explore, at the LO and NLO, both the few- and the many-body sectors;
the description of each system should be consistent with the expected
accuracy order by order irrespectively of the number of particles we
are considering.

The paper is organized as follow. In Section~\ref{sec:LO} we introduce the 
LO potential and apply it to describe atomic $^4$He clusters. Firstly, 
we take into account only the two-body 
force and we explore its predictions in the few-body sector. In addition, we 
give a variational description of the $N\rightarrow\infty$ limit, pointing
out  the necessity of a LO three-body force in order to prevent the collapse 
of the system. Secondly, we explore the predictions with this additional
three-body force and show that the few-body sector  can be described within 
the expected LO accuracy for specific values of the force ranges.

In Section~\ref{sec:NLO} we introduce the NLO potential. Following the same scheme as in
the previous section, we start considering the NLO two-body force. We give a brief
description of the running of the coupling constants and of the energies in the
few-body sector as a function of the two-body range. In addition, we show that
at short ranges the potential develops a repulsive barrier, however in the
continuum limit ($N\rightarrow\infty$) the system turns out to be unstable in any case.  
Then, we proceed with the study of the systems including the LO three-body 
force, we explore the dependence of the few-body binding energies on the 
force ranges selecting those cases in which the description remains
inside the expected accuracy. It should be noticed that for some specific values 
of the two-body range the binding energy of the three-body system is
well reproduced indicating a null contribution of the three-body
force. We show that in those points the continuum system is unstable, suggesting
the existence of a subleading three-body (or eventually four-body)
force. Finally we introduce the NLO three-body force and study the
atom-dimer scattering length together with the binding energies of the
$N\le7$ systems.

In Section~\ref{sec:conclusions} we summarize our findings and outline possible 
future explorations.

\section{Leading order description}~\label{sec:LO}
In order to determine the LO potential we first study the structure of
the two-particle $s$-wave $S$-matrix inside the universal window. It displays 
a two-pole structure
\begin{equation}
    S(k)=\frac{k+i/a_B}{k-i/a_B}\, \frac{k+i/r_B}{k-i/r_B} \, ,
    \label{eq:universalS}
\end{equation}
with the two poles on the imaginary axis,
one fixed by the two-body energy $E_2=-\hbar^2/m a_B^2$,
corresponding to a true bound ($a>0$) or virtual ($a<0$) state, and
the second one at $k
=i/r_B$, with $r_B=a-a_B$, of a spurious character, due to the asymptotic
behavior of the wave
function~\cite{frautschisteven:1963_undefined,newton:1982_}. It can be
shown that $r_B\sim~1/\Lambda$, relating the value of the cut-off to
the second pole. Moreover, the
$S$-matrix of Eq.~(\ref{eq:universalS}) is equivalent to a second
order ERE in which all higher terms are equal to zero
\begin{equation}
  k\cot\delta = -\frac{1}{a} + \frac{1}{2}r_e k^2\,,
  \label{eq:ere}
\end{equation}
with the effective range $r_e$ completely determined by the relation
\begin{equation}
  a r_e = 2 a_B r_B\,.
  \label{eq:ereBis}
\end{equation}
The simplest local potential which reproduces such a basic $S$-matrix
is the Eckart's
potential~\cite{eckart:1930_Phys.Rev.,bargmann:1949_Rev.Mod.Phys.};
however, it has been shown that inside the universal window all the
two-body local potentials are
equivalent~\cite{kievsky:2021_Annu.Rev.Nucl.Part.Sci.}
because the shape parameters, determining the importance of the
successive terms, are very
small~\cite{bethe:1949_Phys.Rev.,blatt:1949_Phys.Rev.}, and the
ERE expression in Eq.~(\ref{eq:ereBis}) is fulfilled up to second
order in the $r_e/a$. This  justifies the use of different forms for a LO potential,
in particular the Gaussian form has been extensively used. In
particular this form was used to characterize the universal window
determining the paths along which very different systems can be
placed~\cite{kievsky:2021_Annu.Rev.Nucl.Part.Sci.,deltuva:2020_Phys.Rev.C,kievsky:2020_Phys.Rev.A}.

In the following we summarize the reference data we are going to use.
They are generated by the LM2M2 helium-helium potential~\cite{aziz:1991_J.Chem.Phys.} for which extremely
accurate numerical results, up to the four-body ground-state energy,
exist~\cite{hiyama:2012_Phys.Rev.Aa}. For the five- and the six-particle ground
state energy we use the results of
Ref.~{\cite{bazak:2020_Phys.Rev.A}. The mass used in all the calculations
is $\hbar^2/m = 43.281307~\text{K}a_0^2$. The two-body scattering length is
$\bar{a}=189.415~a_0$, and the effective range $\bar r_e=13.845~a_0$, resulting
in  the small parameter $\varepsilon=\bar r_e/\bar a \approx 7\%$.  The two-body
ground-state energy fixes  the binding length $\bar a_B=182.221~a_0$ and $\bar r_B=7.194~a_0$.  These
reference data are summarized in Table~\ref{tab:lm2m2Data}.
\begin{table}
  \caption{Reference binding $\bar E_N$ and excited $\bar E^*_N$ energies, in
  mK, for the $^4$He $N$-clusters obtained for the LM2M2
  potential~\cite{aziz:1991_J.Chem.Phys.}. The value of the mass is $\hbar^2/m =
  43.281307~\text{K}a_0^2$. The value of the scattering length is $\bar
  a=189.415~a_0$, of the effective range $\bar r_e=13.845~a_0$, of  $\bar
  a_B=182.221~a_0$, and of $\bar r_B=7.194~a_0$.}
  \label{tab:lm2m2Data}
  \begin{tabular*}{0.7\textwidth} {@{\extracolsep{\fill}}c | l l@{}}
\hline\hline
N & $\bar E_N$(mK) & $\bar E_N^*$(mK)\\
\hline 
2 & -1.30348 \\
3 & -126.40 & -2.2706\\
4 & -558.98~\cite{hiyama:2012_Phys.Rev.Aa} &
-127.33~\cite{hiyama:2012_Phys.Rev.Aa} \\
5 & -1300~\cite{bazak:2020_Phys.Rev.A}  \\
6 & -2315~\cite{bazak:2020_Phys.Rev.A}\\
7 & -3571~\cite{bazak:2020_Phys.Rev.A}\\
\hline
\hline
\end{tabular*}
\end{table}
\subsection{Two-body force}

In the two-body sector, the LO description is given by a Gaussian potential
\begin{equation}
 V_{\text{LO}}(r)= V_0 \,e^{-(r/r_0)^2}\,.
  \label{eq:GaussianLO}
\end{equation}
In the spirit of the EFT approach, the Gaussian range $r_0$ represents the cut-off
of the theory and $V_0$ the low-energy constant that can be fixed by one {\it
experimental} datum, for instance the scattering length $\bar a$.  We can introduce
the dimensionless constant $C_0 = V_0/(\hbar^2/mr_0^2)$ and study its flow as
the cut-off is removed $r_0/\bar a\rightarrow 0$, i.e. in the scaling limit. This
flow is shown in Fig.~\ref{fig:runningC0} together with a quadratic fit. As
already noted in
Refs.~\cite{alvarez-rodriguez:2016_Phys.Rev.A,bazak:2016_Phys.Rev.A,kievsky:2017_Few-BodySyst.},
the scaling limit is well defined $C_0 \rightarrow -2.68402(3)$; moreover, at
fixed $r_0$, this is the value at which the scattering length is infinite. 
\begin{figure}
  \begin{center} 
    \includegraphics[width=0.7\linewidth]{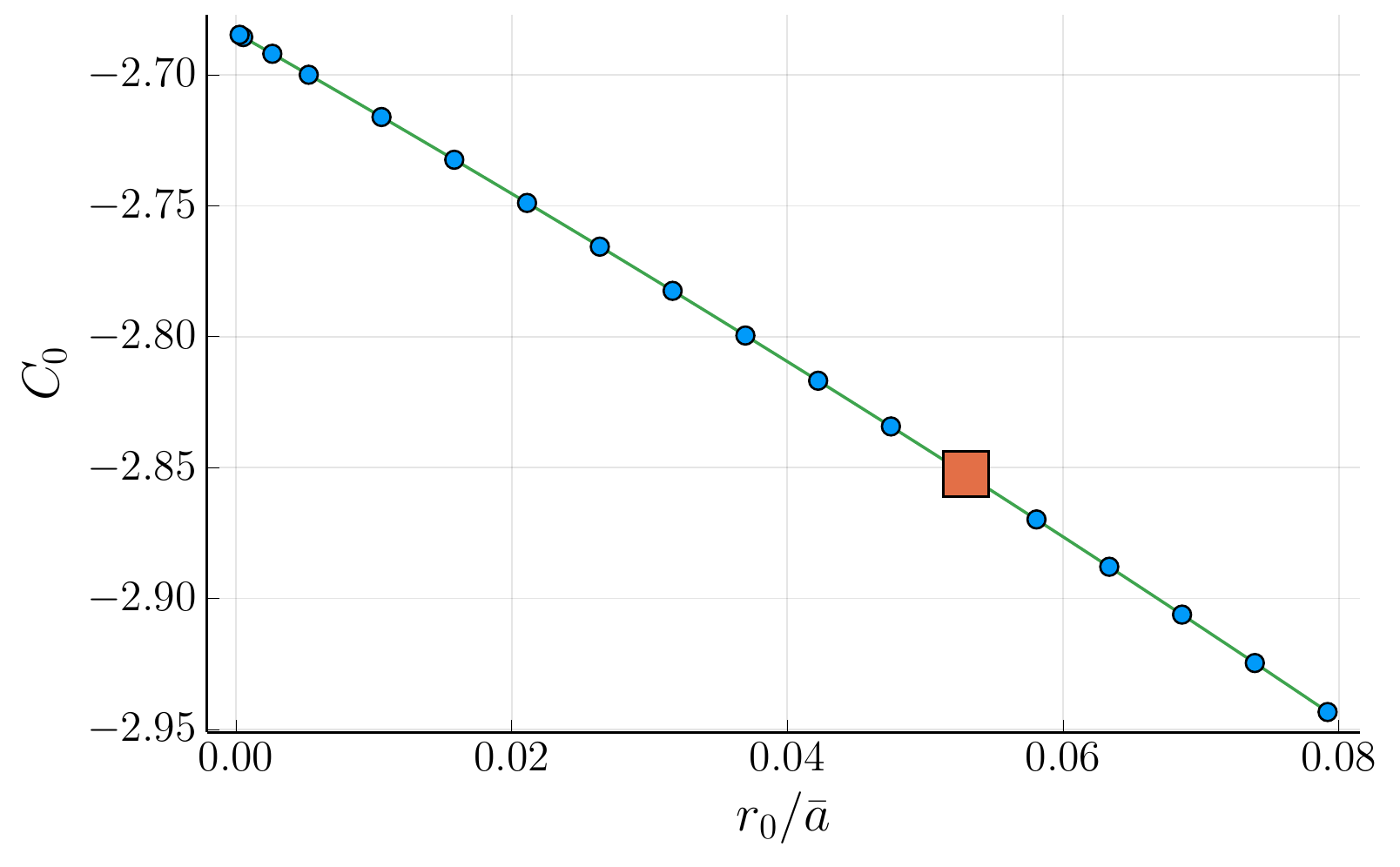}
  \end{center}
  \caption{Running of the dimensionless constant $C_0 = V_0/(\hbar^2/mr_0^2)$ as
  a function of $r_0/\bar a$ at fixed scattering length $\bar
  a=189.415~a_0$. The points correspond to the calculated values, while the
  solid line is a quadratic fit. The limit value 
  for $r_0/\bar a\rightarrow 0$ is $C_0=-2.68402(3)$, and it is
  in agreement with the previous results in
  literature~\cite{alvarez-rodriguez:2016_Phys.Rev.A,bazak:2016_Phys.Rev.A,kievsky:2017_Few-BodySyst.}.
  The square point shows the ratio at which $\bar a_B$ and $\bar r_e$ are well reproduced.}
  \label{fig:runningC0}
\end{figure}
Varying the ratio $r_0/\bar a$, the effective range, the
binding length, and $r_B$ could differ from the LM2M2 values. 
For instance, in the scaling limit 
$a_B\rightarrow \bar a$ with
\begin{equation}
  r_B  \sim 0.7179 \,r_0\,,
  \label{eq:limitAb}
\end{equation}
and,
as anticipated, the second pole of the $S$-matrix, which is proportional 
to $1/r_B$,  is sent to infinity. 
If we want to describe this pole,  both the Gaussian
strength and the range have to be fine tuned
\begin{equation}
  \begin{aligned}
    V_0 &= -1.22717064~\text{K} \\
    r_0 &= 10.03018708~a_0\,,
  \end{aligned}
  \label{eq:valuesLO}
\end{equation}
which  corresponds,in Fig.~\ref{fig:runningC0}, to the square point
at $r_0/\bar a= 5.29535\cdot 10^{-2}$ and $C_0=-2.85248$.
\begin{figure}
  \begin{center} 
    \includegraphics[width=0.7\linewidth]{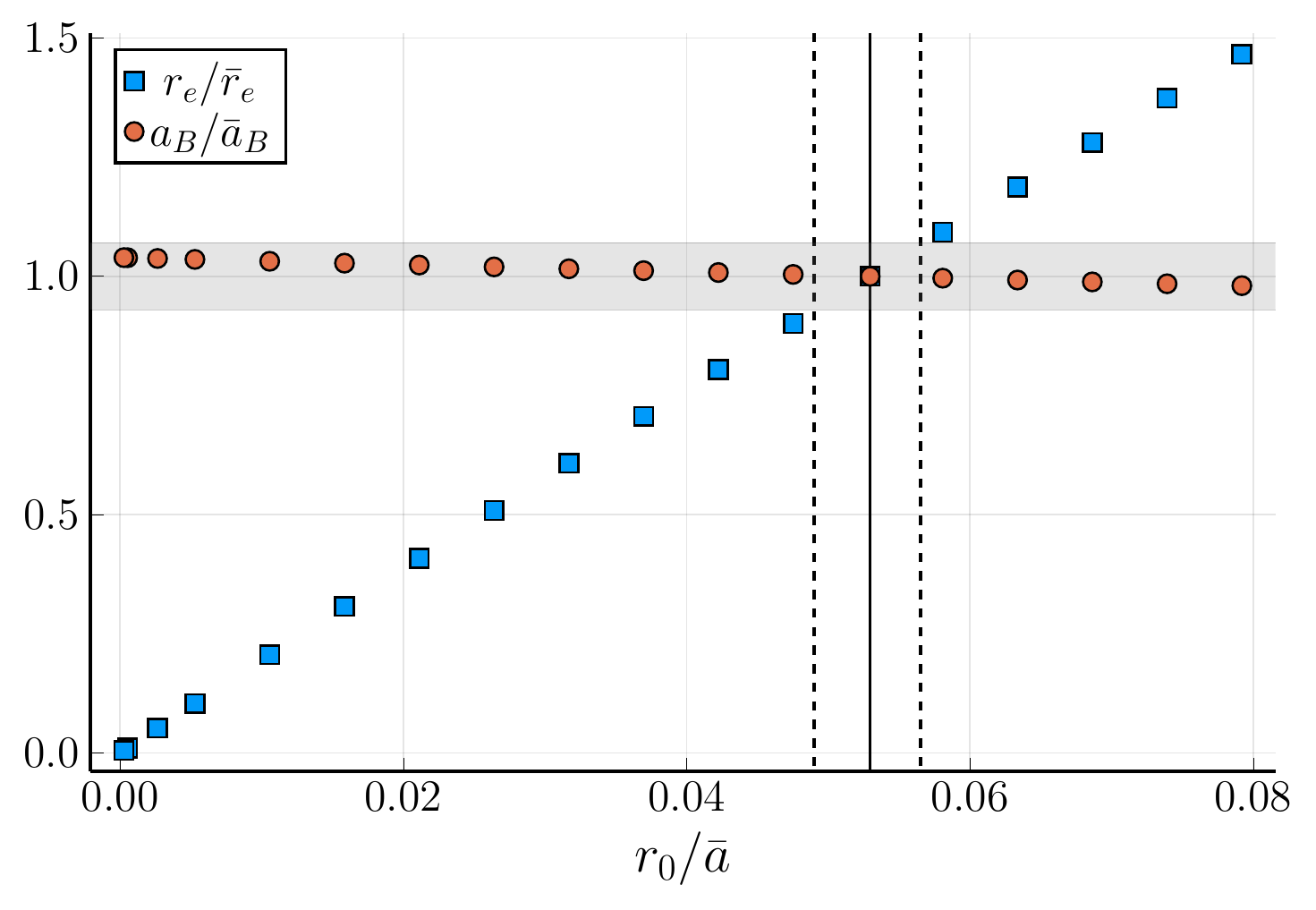}
  \end{center}
  \caption{The ratios $a_B/\bar a_B$ (red circles) and  
  $r_e/\bar r_e $ (blue squares) as a function of the Gaussian range
  $r_0$ (in units of $\bar a $). The gray band represent the
  $\varepsilon=7\%$
  departure from the LM2M value. The binding length $a_B$ is always inside the
  LO-$\varepsilon$ band, while there is only small range of $r_0/\bar a$,
  identified by the two vertical dashed lines, that allows the effective
  range to be inside that band. Inside this range there is a special
  point $r_0/\bar a= 5.29535\cdot 10^{-2}$, identified by the vertical solid
  line, where $\bar a$, $\bar a_B$, and $\bar r_e$ are simultaneously
  reproduced.}
  \label{fig:reabLO}
\end{figure}
Values of $r_e$ and $a_B$ along the flux are shown in 
Fig.~\ref{fig:reabLO}, where the red circles represent the ratio of the 
binding length with respect to the LM2M one, $a_B/\bar a_B$,
while the blue square points represent the ratio of the effective range
with respect to the LM2M one, $r_e/\bar r_e$. 
The gray band indicates the $\varepsilon \approx 7\%$ departure from the LM2M2
value, that one could consider the prediction strip for a LO
description with a small parameter $\varepsilon$. The binding length is 
predicted inside that zone all along the flux, while there is only a small range 
of $r_0/\bar a$  for which also the effective range resides
inside the LO 7\% band. Inside this range, which  is represented by the two vertical-dashed lines in
Fig.~\ref{fig:reabLO}, 
there is the special point 
for which the LO predictions exactly match the LM2M2 values.
\begin{figure}
  \begin{center} 
    \includegraphics[width=0.7\linewidth]{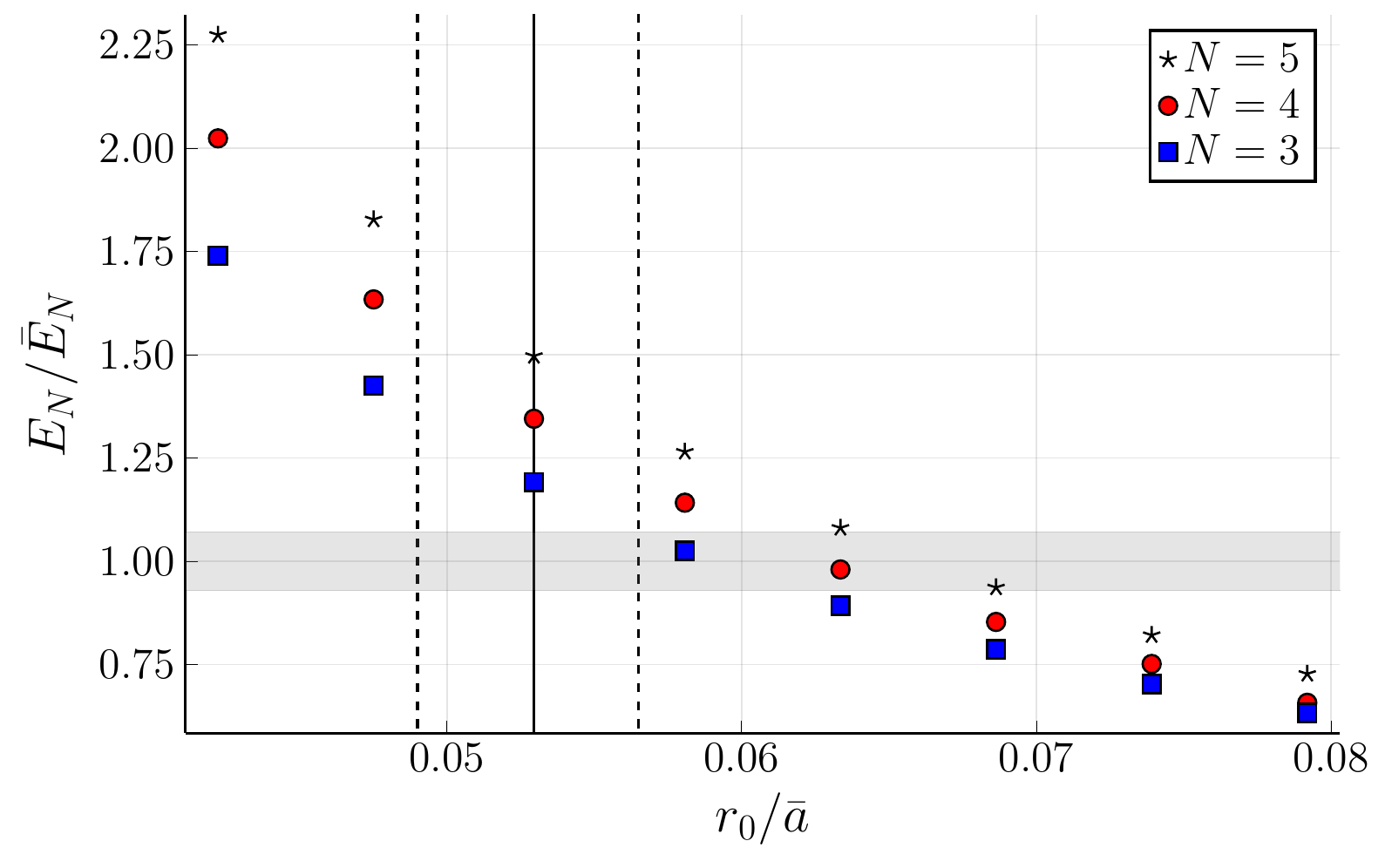}
  \end{center}
  \caption{The ratio of the LO ground-state energy with
  respect to the LM2M2 value for three $E_3/\bar E_3$ (blue squares), four
  $E_4/\bar E_4$ (red circles), and five $E_5/\bar E_5$ (black stars) particle clusters.  The gray band represent the
  $\varepsilon=7\%$ departure from the LM2M value. The vertical lines are the
  same as in Fig.~\ref{fig:reabLO}; the two dashed lines delimit the range where
  the effective range is reproduced within the 7\% and the solid line indicates
  the $r_0/\bar a$ value for which all the two-body LM2M2 data are reproduced.}
  \label{fig:3BLOWo}
\end{figure}

We extend the exploration of the LO description to $N$-body clusters, limiting ourselves to the two-body potential.
In Fig.~\ref{fig:3BLOWo} we show the trend for the three-, four-, and
five-particle ground-state energies scaled with the LM2M2 values
given in Table~\ref{tab:lm2m2Data}. We clearly see that as
$r_0\rightarrow 0$ (scaling limit), $E_N\rightarrow\infty$, because of the well known Thomas
collapse~\cite{thomas:1935_Phys.Rev.}. We also note that even if we
fine-tune the value of the cut-off $r_0$ inside the range marked by
the two vertical-dashed lines, where we
reproduce the two-body observables, the three- and four-body data are
not reproduced within the LO. 
Moreover, we observe that the distance
from the LM2M2 value grows as a function of the size of the cluster
even if we fix the value of the cut-off. In fact, in the 
thermodynamical limit, $N\rightarrow\infty$ there is a collapse
of the system for all finite values of the Gaussian range $r_0$.
We can use the Hyperspherical Harmonics $K=0$
approximation to give a variational bound to the energy of the clusters 
in the $N\rightarrow \infty$ limit~\footnote{M. Gattobigio, and A. Kievsky, in preparation}; 
the ground-state energy per particle is bounded from above by 
\begin{equation}
  \frac{E_N}{N} =  \frac{V_0}{2} N \,,
  \label{eq:collapseLO}
\end{equation}
showing that in this limit the system is unstable.
This can be taken as a complementary evidence that in the LO
description, even at finite cut-off, the theory needs a three-body
force to stabilize the continuum limit of the system. 

\subsection{Three-body force}
From the above discussion we have observed that using a Gaussian
potential to describe the LO and taking the limit $r_0/a\rightarrow 0$,
the three-body ground-state energy $E_3$ diverges as $ 1/r_0^2$.
Furthermore, if the Gaussian range is fixed to describe the two-pole
structure of the $S$-matrix choosing the values given in Eq.~(\ref{eq:valuesLO}), 
we still observe the collapse given by Eq.~(\ref{eq:collapseLO})
as $N\rightarrow\infty$.
On the other hand we would like to see, using the inspired EFT
potential at LO, all the particle sectors up to the continuum limit 
described inside the LO prediction. 
For instance, at the point fixed by Eq.~(\ref{eq:valuesLO}) even the 
three-body bound state, $E_3=-150.57$~mK, 
is outside the LO band, as one can see in Fig.~\ref{fig:3BLOWo}.
Following the EFT prescription we include a three-body force at LO
and proceed with computation of the ground state energies of the
$N$-body clusters.
The three-body force is chosen of the following from
\begin{equation}
  W_{\text{LO}} = W_0\,e^{-(r_{12}^2 + r_{13}^2 + r_{23}^2)/\rho_0^2}\,,
  \label{eq:3BLO}
\end{equation}
with $r_{ij}$ the distance between particle $i$ and particle $j$,
$W_0$ the strength,  and $\rho_0$ the range of the force. The force is
determined by one three-body datum, to this purpose we use the $E_3$
ground-state energy obtained with the LM2M2. It should be noticed
that in EFT the range $\rho_0$ is sent to zero together with $r_0$.  
In the following we analyze the dependence of $E_N$ on different
choices of $r_0$ and $\rho_0$. The resulting limit cycle in the renormalization group flow  has been recently
studied~\cite{recchia:2022_Few-BodySyst}. Here
we determine different combinations of $(V_0,r_0)$ values, all of them
reproducing the two-body scattering length $\bar a$. For each pair
we associate a family of pair values $(W_0,\rho_0)$ leading to the
same  three-body binding energy $E_3$.
\begin{figure}
  \begin{center} 
    \includegraphics[width=0.7\linewidth]{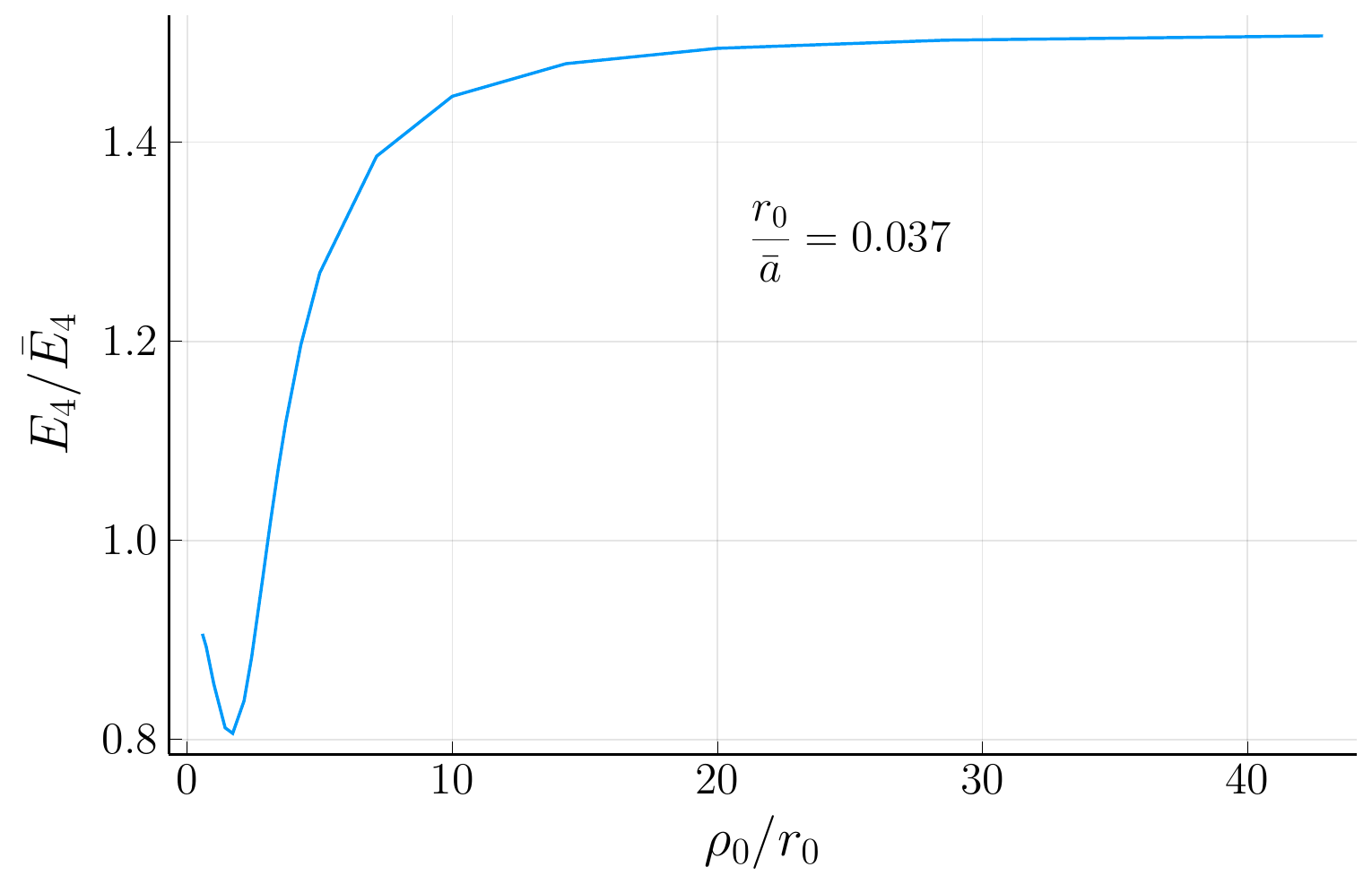}%
  \end{center}
  \caption{The ratio $E_4/\bar E_4$ as a function of $\rho_0/r_0$. 
  In the figure, the specific case $r_0/\bar a=0.037$ 
  is shown. For each value of $\rho_0$, the three-body 
  strength $W_0$ has been fixed in order to verify $E_3/\bar E_3=1$.
  The ratio $E_4/\bar E_4$ varies between two values, and if the 
  minimum is below one, there is a special 
  value of the pair $(W_0,\rho_0)$ for which $E_4/\bar E_4=1$.}
  \label{fig:trendLO}
\end{figure}
We would like to show that inside these families of pairs there is a best choice 
which allows for the optimum description of the multi-particle sectors, starting from
the four-particle
one~\cite{gattobigio:2011_Phys.Rev.A,kievsky:2020_Phys.Rev.A}.
This can be seen in Fig.~\ref{fig:trendLO}, where the ratio 
$E_4/\bar E_4$ is calculated as a function of $\rho_0/r_0$; the ratio 
has a bell-type shape and it varies between a minimum, which is
usually attained for $\rho_0/r_0 \sim 1$, and a maximum in the limit
of $\rho_0/r_0 \rightarrow +\infty$, where the three-body force tends 
to be a constant (the difference between $\bar E_3$ and the three-body
ground-state energy without the three-body force).
In the figure we have selected the value $r_0/\bar a=0.037$, however
this behavior is similar for different ratios and for different particle sectors. 
\begin{figure}
  \begin{center} 
    \includegraphics[width=0.5\linewidth]{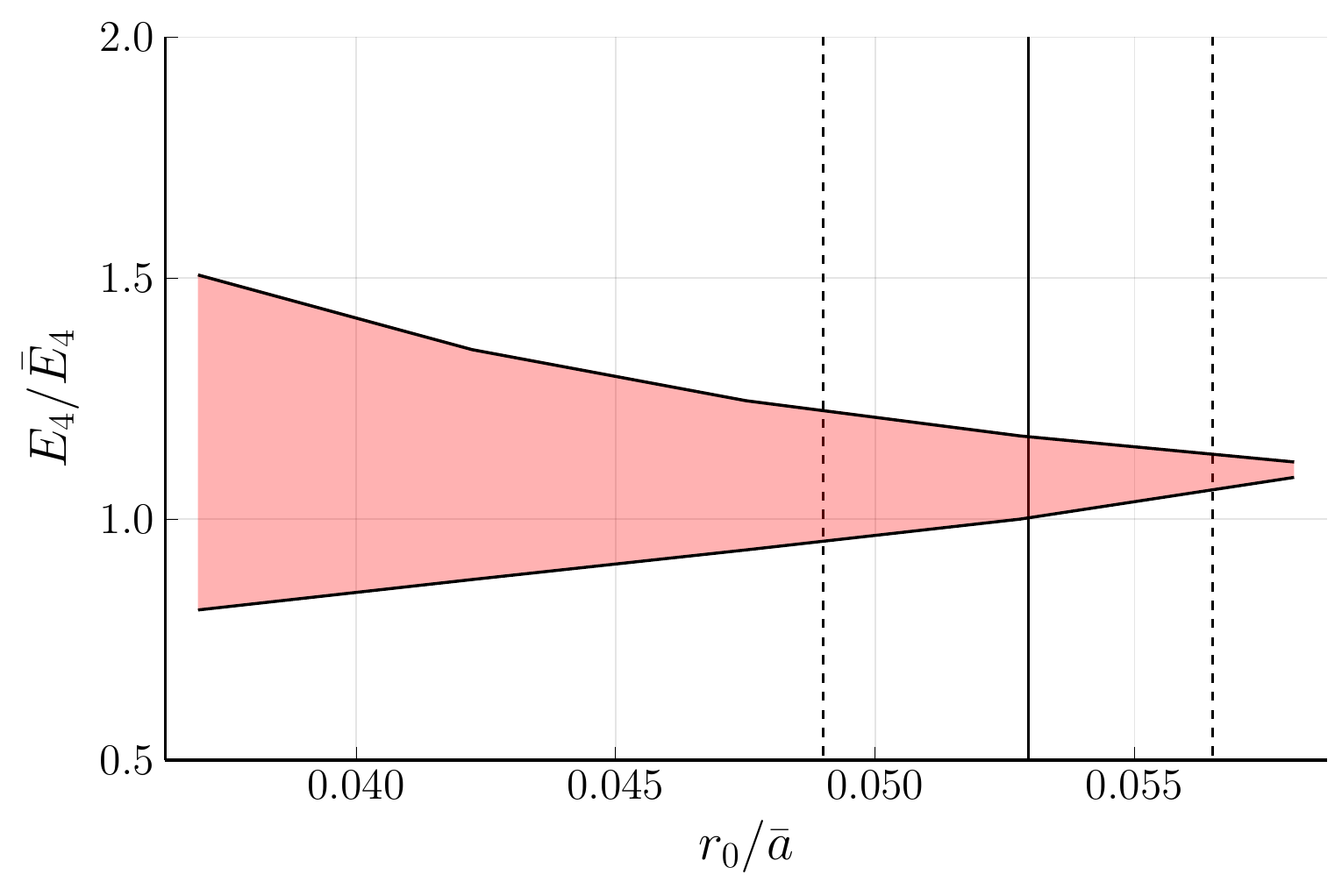}%
    \includegraphics[width=0.5\linewidth]{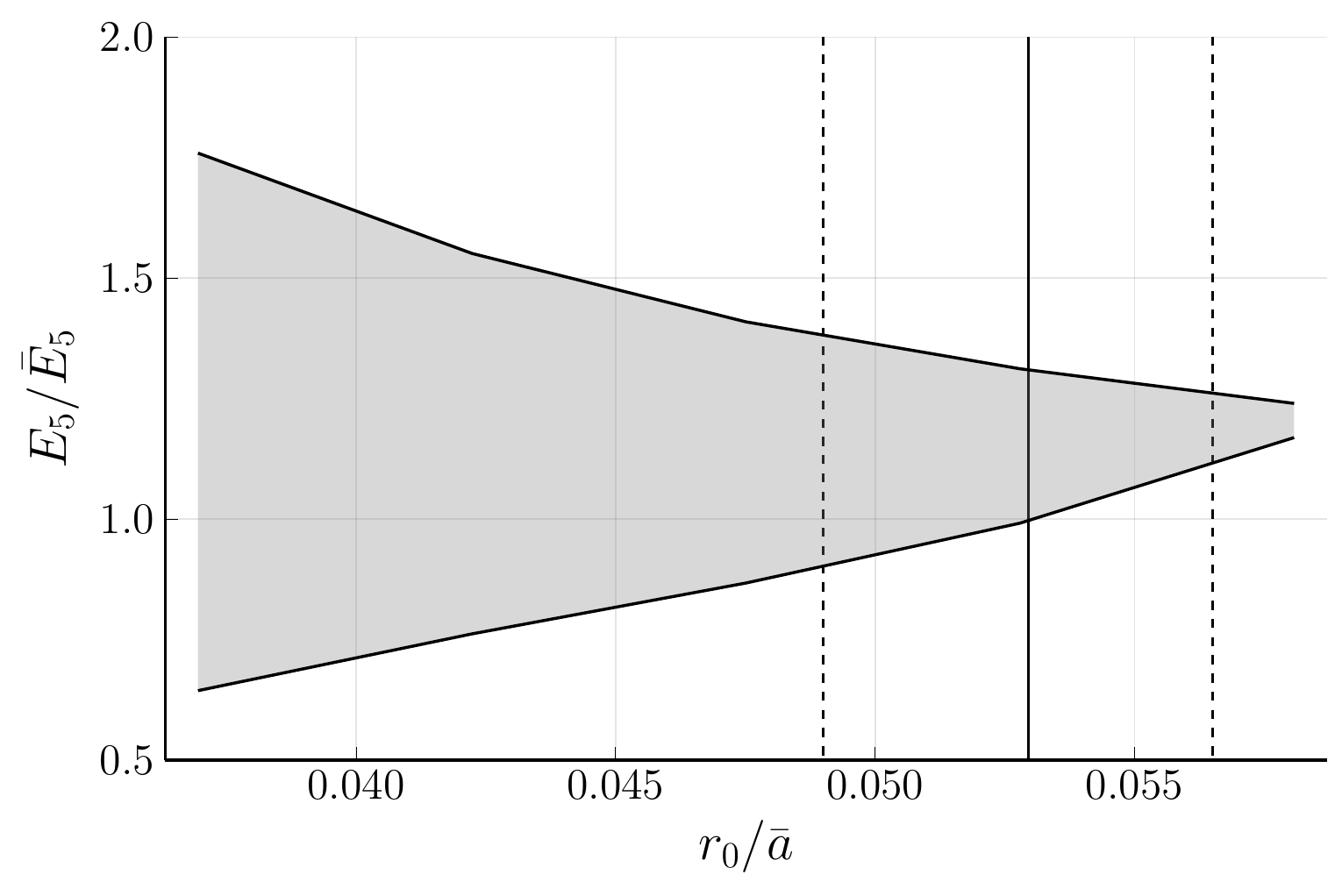}
  \end{center}
  \caption{Range of variability of the ratios
  $E_4/\bar E_4$ (left panel) and $E_5/\bar E_5$ (right panel)
  as a function of the two-body range $r_0/\bar a$. For a fixed
  value of $r_0/\bar a$, the range of variability is given by the maximum and the
  minimum of the curve as given in Fig.~\ref{fig:trendLO} for one case. The vertical
  lines have the same meaning as in Fig.~\ref{fig:reabLO}. For 
  $r_0/\bar a \lesssim 0.053$, on the left of the
  solid-vertical line, it is possible to fix 
  either $E_4/\bar E_4=1$ or $E_5/\bar E_5=1$ with a suitable choice
  of the $(W_0,\rho_0)$ pair. }
  \label{fig:45BLO}
\end{figure}

In Fig.~\ref{fig:45BLO} we report the bands inside which the ratios 
$E_4/\bar E_4$ and $E_5/\bar E_5$ of four and five particles respectively, 
can be found as the ratio $\rho_0/r_0$ varies from its lower to its maximum value.
We observe that for $r_0/\bar a$ values at the left of the special point given in
Eq.~(\ref{eq:valuesLO}), and represented 
in Fig.~\ref{fig:45BLO} by the solid-vertical line,
it is possible to fix the pair
$(W_0,\rho_0)$ in order to reproduce either the four- or the five-body 
ground-state energy. 

To analyze further this fact,
we tune the pairs $(W_0,\rho_0)$ to have $E_4/\bar E_4=1$
whenever is possible, namely for $r_0/\bar a \lesssim 0.053$. In the other cases the ratio
$E_4/\bar E_4$ is set as close as possible to 1. With this prescription we 
calculate the ground-state energy for $N=5,6,7$ clusters and the 
results are shown in Fig.~\ref{fig:allLOWith}. 
Analysing the figure from the left to the right,
we observe that the few-body ground-state energies tends to converge to the LM2M2 data
as we move toward the special value of $r_0/\bar a$. Around this point,
the predicted energies are well inside the $\varepsilon=7\%$ 
deviation from LM2M2 data maintaining the LO accuracy. This point has been already noticed in 
Ref.~\cite{kievsky:2020_Phys.Rev.A} where a different 
helium potential has been used as reference. Moreover, in 
Ref.~\cite{kievsky:2020_Phys.Rev.A} it has been shown that also 
the saturation energy is predicted within the LO uncertainty. 
The present analysis extends these findings showing
that there is a small range of values of the cut-offs, $r_0$ and
$\rho_0$, allowing for a LO accuracy of the energy per particle $E_N/N$.
\begin{figure}
  \begin{center} 
    \includegraphics[width=0.7\linewidth]{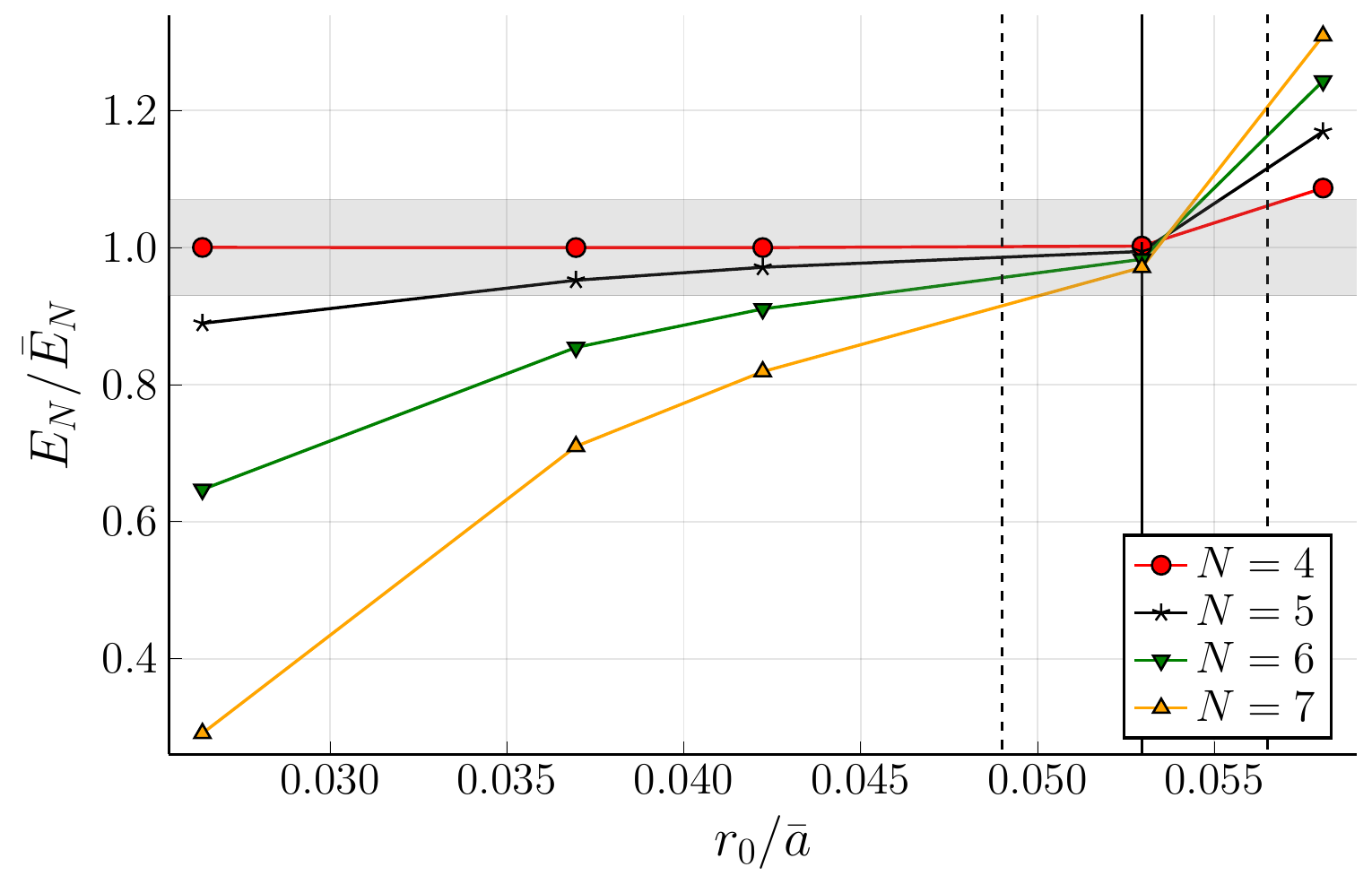}%
  \end{center}
  \caption{The ratios $E_N/\bar E_N$ for $N=4,5,6,7$ are given as a
  function of $r_0/\bar a$. The three-body force has been fixed by
  imposing $E_4/\bar E_4$ as close as possible to 1. The vertical
  lines are the same as in Fig.~\ref{fig:reabLO}, and the gray strip
  corresponds to the $\varepsilon=7\%$ departure from LM2M2 data.
  When the four-body energy $E_4=\bar E_4$, the five- and six-body
  clusters are found to be less bound than the LM2M2 data, with the
  best description at the point given by Eq.~(\ref{eq:valuesLO}), i.e.
  where the second pole of the $S$-matrix is well described.  In this
  point the predictions are all inside the $\varepsilon$-LO strip.
  For $r_0/\bar a > 0.053$ the opposite trend is verified.}
  \label{fig:allLOWith}
\end{figure}

\section{Next-to-leading order description}~\label{sec:NLO}
In this section we introduce the next-to-leading order (NLO) term
of the EFT inspired potential. At this order we expect to increase the accuracy 
of the description at the level of $\varepsilon^2\approx 5\permil$. 


\subsection{Two-body force}

Following the same counting criterion of EFT~\cite{epelbaum:2017_Nucl.Phys.B},
the two-body NLO potential is
\begin{equation}
  V_{\text{NLO}}(r)= V_0\, e^{-(r/r_0)^2} + V_1\, \frac{r^2}{r_0^2} e^{-(r/r_0)^2}\,,
  \label{eq:nlo}
\end{equation}
where the additional term is proportional to the square-particle
distance. In the following the range of the two Gaussian functions are
kept equal, but clearly this is not the
only possible choice~\cite{epelbaum:2018_Commun.Theor.Phys.}.

\begin{figure}
  \begin{center} 
    \includegraphics[width=0.7\linewidth]{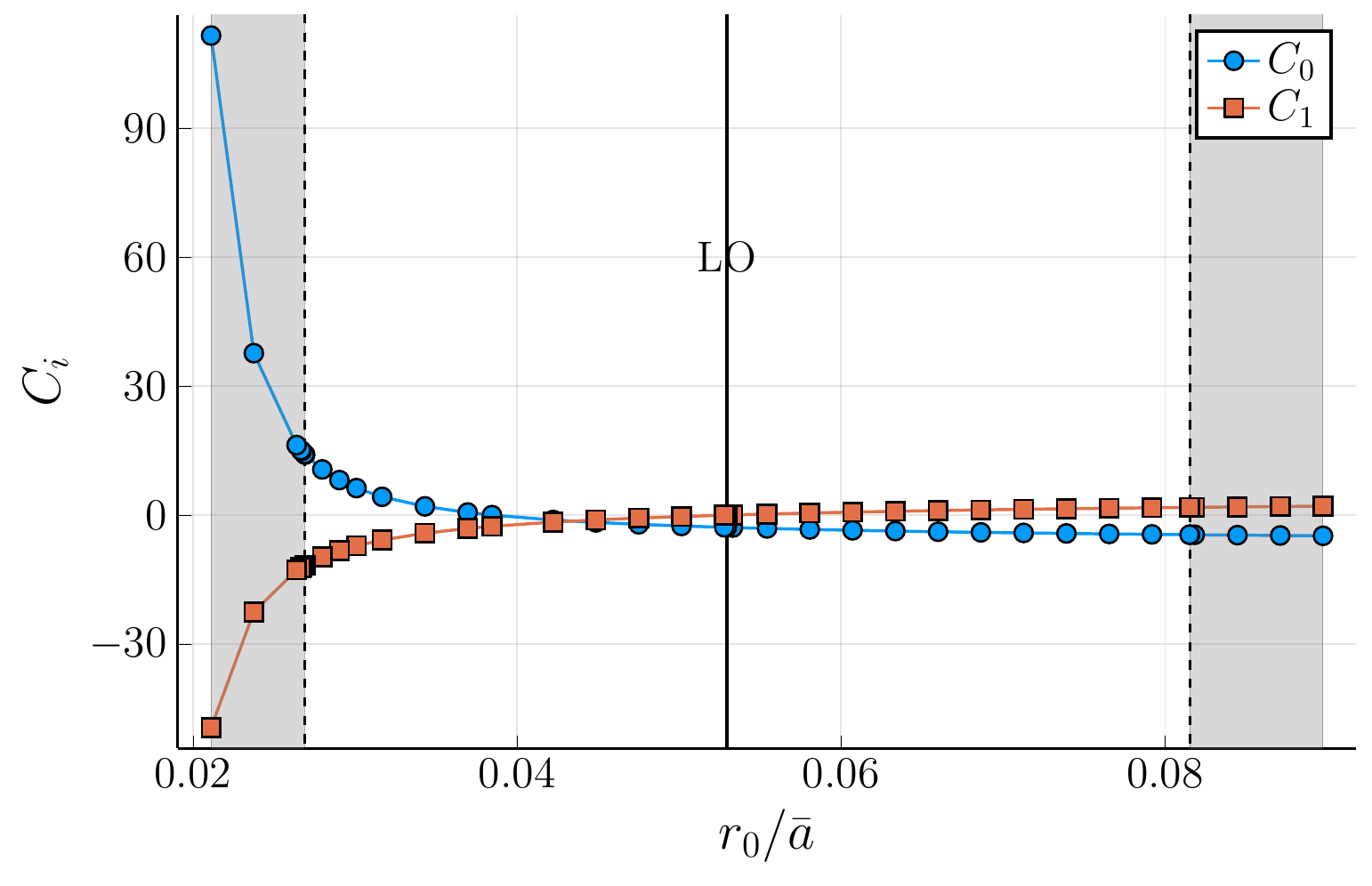}
  \end{center}
  \caption{Running of the dimensionless constants $C_0 =
  V_0/(\hbar^2/mr_0^2)$ and $C_1 = V_1/(\hbar^2/mr_0^2)$ as a function of
  $r_0/\bar a$ at fixed scattering length $\bar a=189.415~a_0$ and 
  effective range $\bar r_e=13.845~a_0$. The gray zones 
  correspond to $E_3/\bar E_3 < 1$. 
  The vertical-dashed lines indicate the two values of $r_0/\bar a$ 
  where  $E_3/\bar E_3 = 1$.
  The center-dashed line shows the $C_1=0$ case, 
  where the LO and the NLO potentials concide.}
  \label{fig:runningNLO}
\end{figure}
For different choices of the range $r_0$, the NLO potential has two
LECS, $V_0$ and $V_1$. There
is a whole family of strength values which allows to reproduce the structure of
the $S$-matrix given in Eq.~(\ref{eq:universalS}). Introducing the dimensionless 
strengths $C_0=V_0/(\hbar^2/mr_0^2)$ and $C_1=V_1/(\hbar^2/mr_0^2)$,
we fix them as a function of the range $r_0/\bar a$ to reproduce both the
scattering length $\bar a$ and the effective range $\bar r_e$.  
In Fig.~\ref{fig:runningNLO} we trace the two strengths (or LECs), $C_0$ and 
$C_1$ as a function of $r_0/\bar a$.
We observe that there is a special point, indicated by the 
vertical-solid line (labeled LO), where $C_1=0$ and the NLO description coincides
with the LO description of Eq.~(\ref{eq:valuesLO}). In addition, there
are two special points, marked by the two vertical-dashed lines, 
where $E_3/\bar E_3=1$. In this case the three-body force should give
a null contribution to the three-body ground state. 
Beyond these lines, we have the zone where $E_3/\bar E_3>1$ indicated 
by the gray strips, for which an attractive three-body force is needed. 
The scaling limit of $C_0$ and
$C_1$ is not finite; both LEC's have an essential singularity  at $r_0=0$, $C_i
\sim \exp[\alpha_i (\bar a/r_0) + \beta_i(\bar a/r_0)^2]$, as can be clearly seen
from Fig.~\ref{fig:logC} ($\alpha_i$ and $\beta_i$ are fitting constants).
\begin{figure}
  \begin{center} 
    \includegraphics[width=0.7\linewidth]{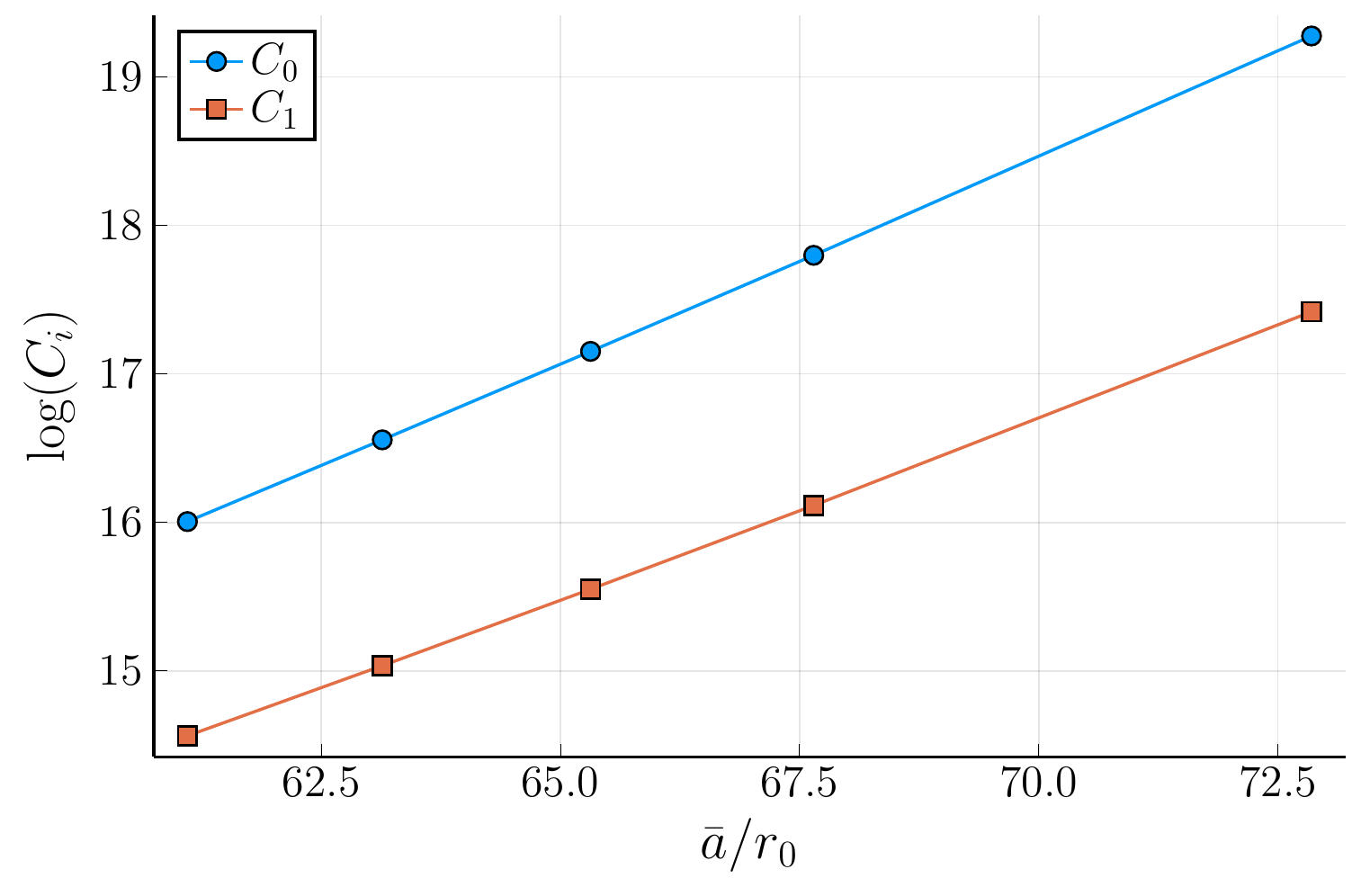}
  \end{center}
  \caption{The log-behavior of the LECs as a function of $\bar a/r_0$ 
  for small values of $r_0$. The two curves
  can be fitted using a polynomial of degree two, showing that $C_i \sim
  \exp(\alpha_i (\bar a/r_0) + \beta_i(\bar a/r_0)^2$. }
  \label{fig:logC}
\end{figure}

The family of potentials having the same scattering length and the same effective
range, but different range, are not phase equivalents; they develop
different shape parameters. In Fig.~\ref{fig:potenziali}
the two-body potential is plotted for three different values of $r_0$,
corresponding to the cases where they give $E_3/\bar E_3=1$ (left and
right plot), and to the special LO point of Eq.~(\ref{eq:valuesLO}) (center
plot) which is also NLO. Interestingly, as $r_0/\bar a\rightarrow 0$,
the potential develops a repulsive core which mimics the LM2M2 one,
however this feature is not enough to prevent the collapse of the
$N\rightarrow\infty$ system. This behavior is illustrated in
Fig.~\ref{fig:NLOWo} where we show $E_N/\bar E_N$ for $N=3,4,5,6$.
As the number of particles grows, the ratio $E_N/\bar E_N$ increases
indicating the instability in the $N\rightarrow\infty$ limit. In
particular this is the case for the lowest value analyzed, $r_0/ a_0= 5.093$,
corresponding to the case $E_3/\bar E_3=1$.  
This instability can be further analyzed using the HH $K=0$ approximation
which gives the asymptotic variational estimate of the 
energy per particle as proportional to the number of particles~\cite{Note1},
\begin{equation}
  \frac{E_N}{N} \propto  N \,.
\end{equation}

\begin{figure}
  \begin{center} 
    \includegraphics[width=0.7\linewidth]{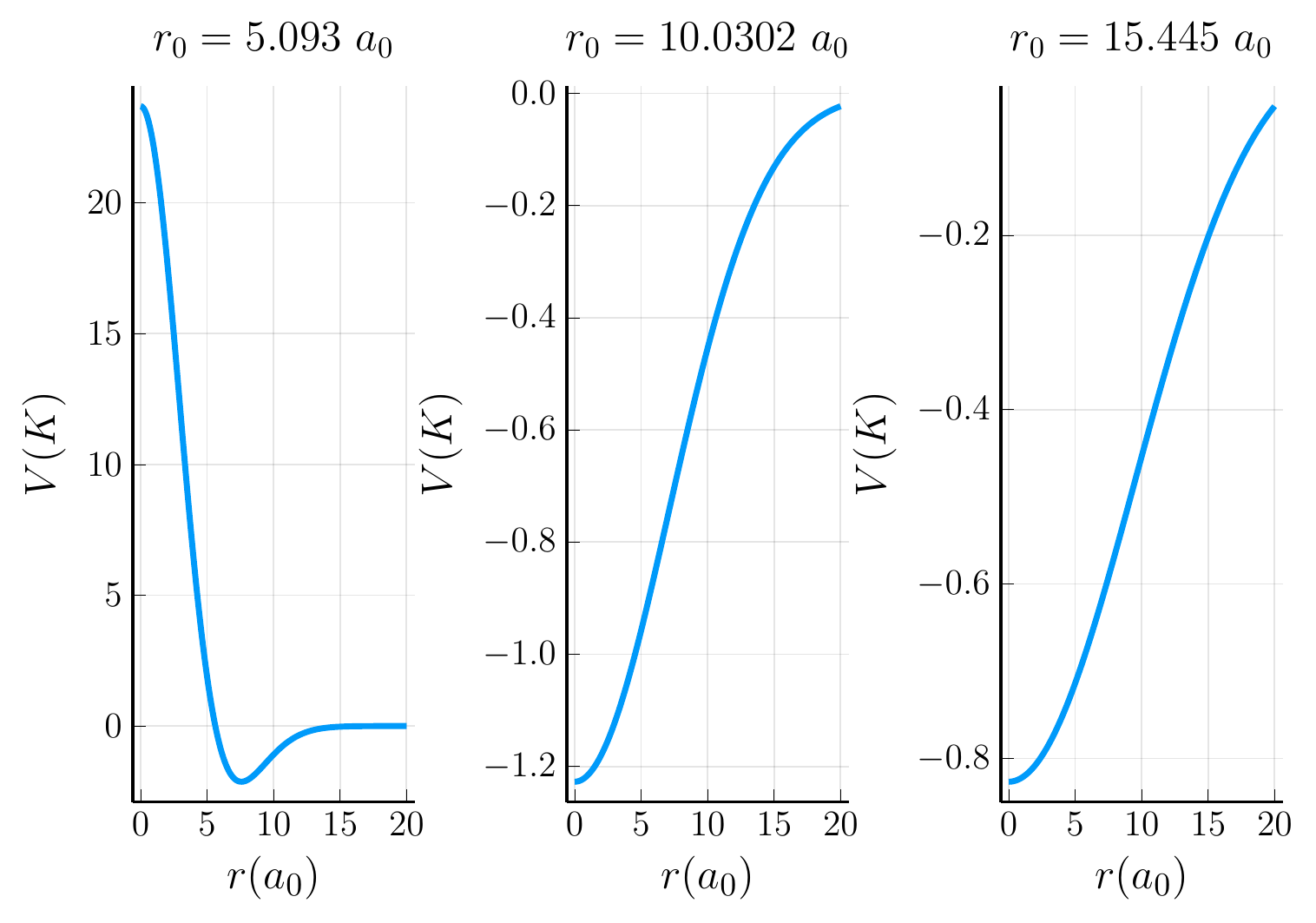}
  \end{center}
  \caption{Profile of the NLO potentials for the three different values of $r_0$. The left
  and right panels correspond to the two extreme cases where the NLO potential
  reproduce the three-body ground-state energy. These two points are
  indicated by the vertical-dashed line in Fig.~\ref{fig:runningNLO}.
  The central panel corresponds 
  to the NLO-LO case (vertical-solid line in Fig.~\ref{fig:runningNLO}).}
  \label{fig:potenziali}
\end{figure}

\begin{figure}
  \begin{center} 
    \includegraphics[width=0.7\linewidth]{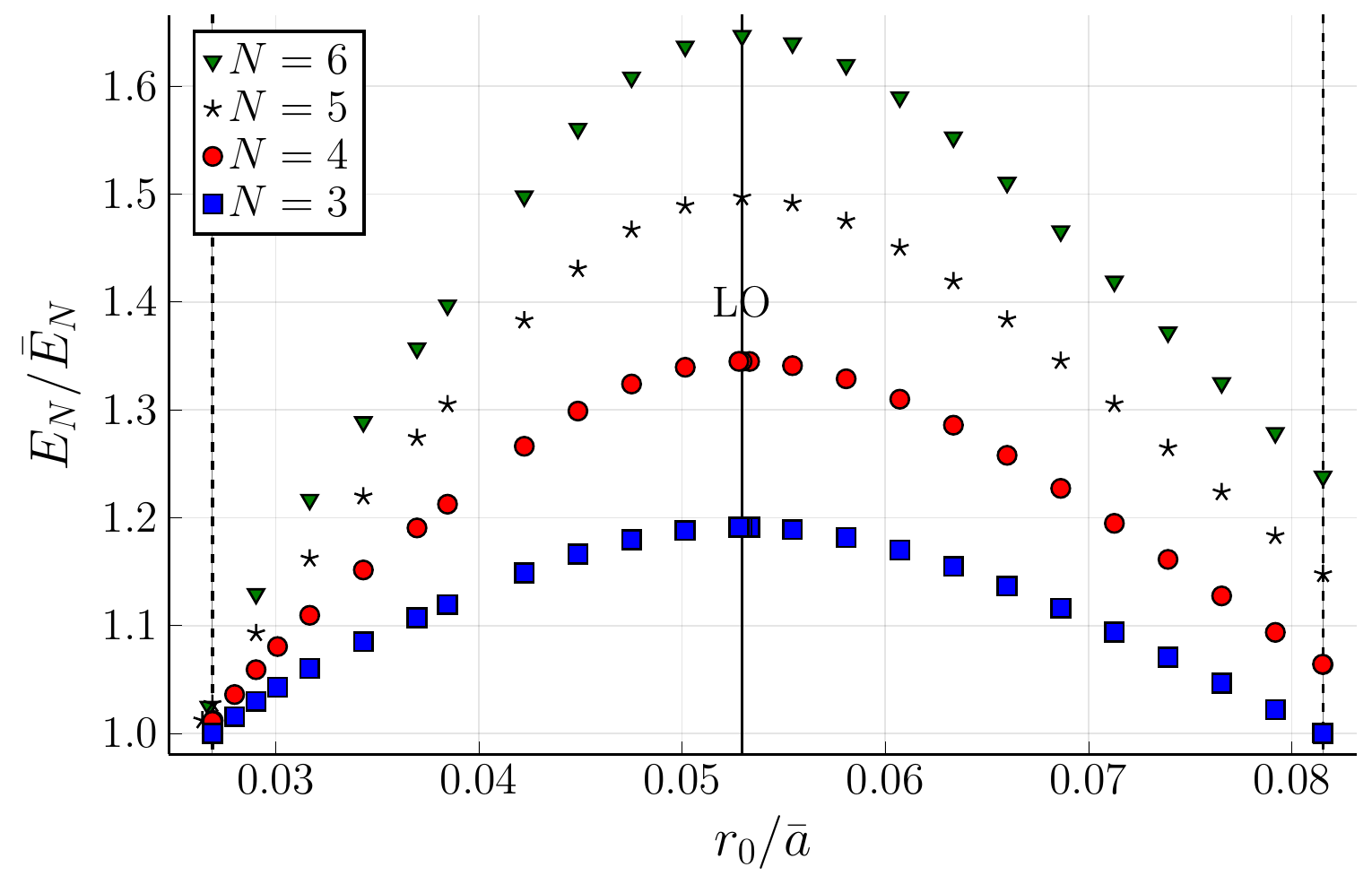}%
  \end{center}
  \caption{The ground-state energy $E_N$ (in units of $\bar E_N$).
  computed with the NLO two-body force of Eq.~(\ref{eq:nlo}).
  The vertical lines are the same as in Fig.~\ref{fig:runningNLO}.}
  \label{fig:NLOWo}
\end{figure}

\subsection{Three-body force}
Contrary to the LO case, the three-body ground state calculated with two-body 
NLO potential does not collapse as $r_0\rightarrow 0$. Approaching this limit
the two-body potential develops a repulsive core, a consequence of the
two parameters, the scattering length and the effective range, it has
to reproduce. For $r_0/a$ values inside the region verifying $E_3/\bar E_3\ge 1$, 
the deepest energy corresponds to the particular case in which $V_1=0$
and the LO and NLO potentials coincide. This is true also for $N>3$ as
shown in Fig.~\ref{fig:NLOWo} where the binding energies for
$N=3,4,5,6,7$ are shown, in units of $\bar E_N$, as functions of the
Gaussian range $r_0$, in units of $\bar a$. A demonstration of why the deepest
energy is reach in that particular point is given in the Appendix~A.

We proceed the study by considering the two-body NLO potential plus
the LO three-body force of Eq.~(\ref{eq:3BLO}). To this end we compute pairs
of $(W_0,\rho_0)$ values from the condition $E_3/\bar E_3=1$.
We select the best pair of values by analyzing the
four-body ground-state energy and, using this
choice, we calculate the $N=5,6,7$ binding energies. 
The results are reported in Fig.~\ref{fig:allNLOWith}, where 
we show the ratios $E_N/\bar E_N$ for $N=4,5,6,7$ as functions of
$r_0/\bar a$ inside the region in which the three-body force is
repulsive. Firstly we observe that, although $E_4$ is very close to the exact value, it is not 
possible to set $E_4/\bar E_4=1$; the best value is obtained 
for the NLO-LO point given by Eq.~(\ref{eq:valuesLO}) and it is inside
the $\varepsilon^2=5\permil$ deviation from the LM2M2 value. The
$E_4/\bar E_4$ ratio remains close to 1 inside the region between the NLO-LO point 
and the lower value of $r_0/\bar a$ for which the three-body binding
energy is well reproduced by solely the NLO two-body potential.
In between these points, there is a special value of $r_0/\bar a
\approx 0.042 $  (and $\rho_0/r_0 \approx 7/8$) such that all the $N$-body systems (at least up to
seven particles) have a ground-state energy which is inside the 
NL0-$\varepsilon^2$ strip. There is a similar point for the higher 
value $r_0/\bar a\approx 0.063$  (and $\rho_0/r_0 \approx 7/12$).
At these two points, the binding energies, $E_N$ with
$N\le 7$ are predicted inside the $5\permil$ strip.

Now we look at the two points, $r_0/\bar a=0.027$ and $0.081$, characterized by the
fact that the three-body binding energy is well reproduced by the 
two-body NLO potential. Accordingly the contribution of the three-body
force, in the three-body system, should be zero. However, if we consider 
only the two-body NLO potential and use
the HH $K=0$ approach to estimate the limit $N\rightarrow\infty$ we
can show that the system is unstable~\cite{Note1}. Therefore we
conclude that the three-body force should have at least two terms that
compensate each other to give zero contribution in the three-body
system but non-zero ones for $N>3$ and stabilizes the systems as $N$ increases. 
To analyse this behavior we introduce the following NLO three-body force 
\begin{equation}
 W_\text{NLO} = W_0 e^{-r_{123}^2/\rho_0^2} +
            W_1 \left(\frac{r_{123}}{\rho_0}\right)^2 e^{-r_{123}^2/\rho_0^2} \,,
  \label{eq:3BNLO}
\end{equation}
where $r_{123}^2 = r_{12}^2+r_{13}^2+r_{23}^2$. To be noticed that
in a perturbative scheme the $W_1$ term can be absorbed 
by the LO term~\cite{hammer:2001_Phys.Lett.B,bedaque:2002_Annu.Rev.Nucl.Part.Sci.}.
In this case, and as discussed in Ref.~\cite{bazak:2019_Phys.Rev.Lett.}, 
a subleading four-body interaction could be introduced. Using a non-perturbative
scheme, the two LECs $W_0$ and $W_1$ are independent.
In order to determine possible values of the additional LEC, $W_1$, we
compute pairs of $W_0,W_1$  values obtained through the 
condition $E_3=\bar E_3$. Next, we study the effects of the different
pairs in the binding energies $E_N$ with $N\le 7$.

We now extend the study to consider the complete NLO potential
consisting in a two plus a three-body term. The results are shown in
Fig.~\ref{fig:NLOfull}, where the ratios $E_N/\bar E_N$ and $a_2/\bar
a_2$ are given as a function of the gaussian range $r_0$, in units of
$\bar a$. From the figure we can conclude that the NLO force is 
sufficiently flexible to describe accurately the $E_N$ binding
energies inside the region in which the three-body force has an
overall repulsive contribution including the point in which it has a
null contribution to the three-body system. All the binding energies
fall inside (or very close to) the $5\permil$ strip. 
To obtain these results the three-body range $\rho_0$ has been varied
and, for each value of $r_0$, the selected value is the one that gives 
the best results. This fact implies that the two
ranges, $r_0$ and $\rho_0$, are correlated and need to be tuned
simultaneously to optimize the capability of the force to
describe systems having different values of $N$. For each value
of $r_0$ there is a particular value of $\rho_0$ that produce the
best results for the binding energies. We infer that this value is
not universal, it takes into account the short range characteristics
of the systems trying to adapt the (repulsive) three-body force
in order to reproduce the packing of the particles as $N$ increases.
It seems that this behavior, for one specific value of $r_0$,
can be codified in one particular value of $\rho_0$.

Finally we incorporate in the study the atom-dimer scattering 
length $a_2$. This observable will give additional information on the
capability of the NLO potential to describe the few-body dynamics.
In Fig.~\ref{fig:a2NLO} we report a preliminary analysis of 
the ratio $a_2/\bar a_2$, where $\bar a_2=218.0~a_0$ is the LM2M2
atom-dimer scattering length. In the figure we show the results
including the two-body NLO force (green symbols), when the LO
three-body force is also included (blue symbols) and when NLO
three-body is considered as well (red symbols).
In the first case we see that there is a parabolic behavior with the
minimum just in the LO/NLO point corresponding to $r_0/\bar a=0.053$.
This behavior is maintained when the LO three-body force is included
though with less variation in all the region analyzed. When the
NLO three-body force is included the observable is well inside
the expected error band.

%
%
%

\begin{figure}
  \begin{center} 
    \includegraphics[width=0.7\linewidth]{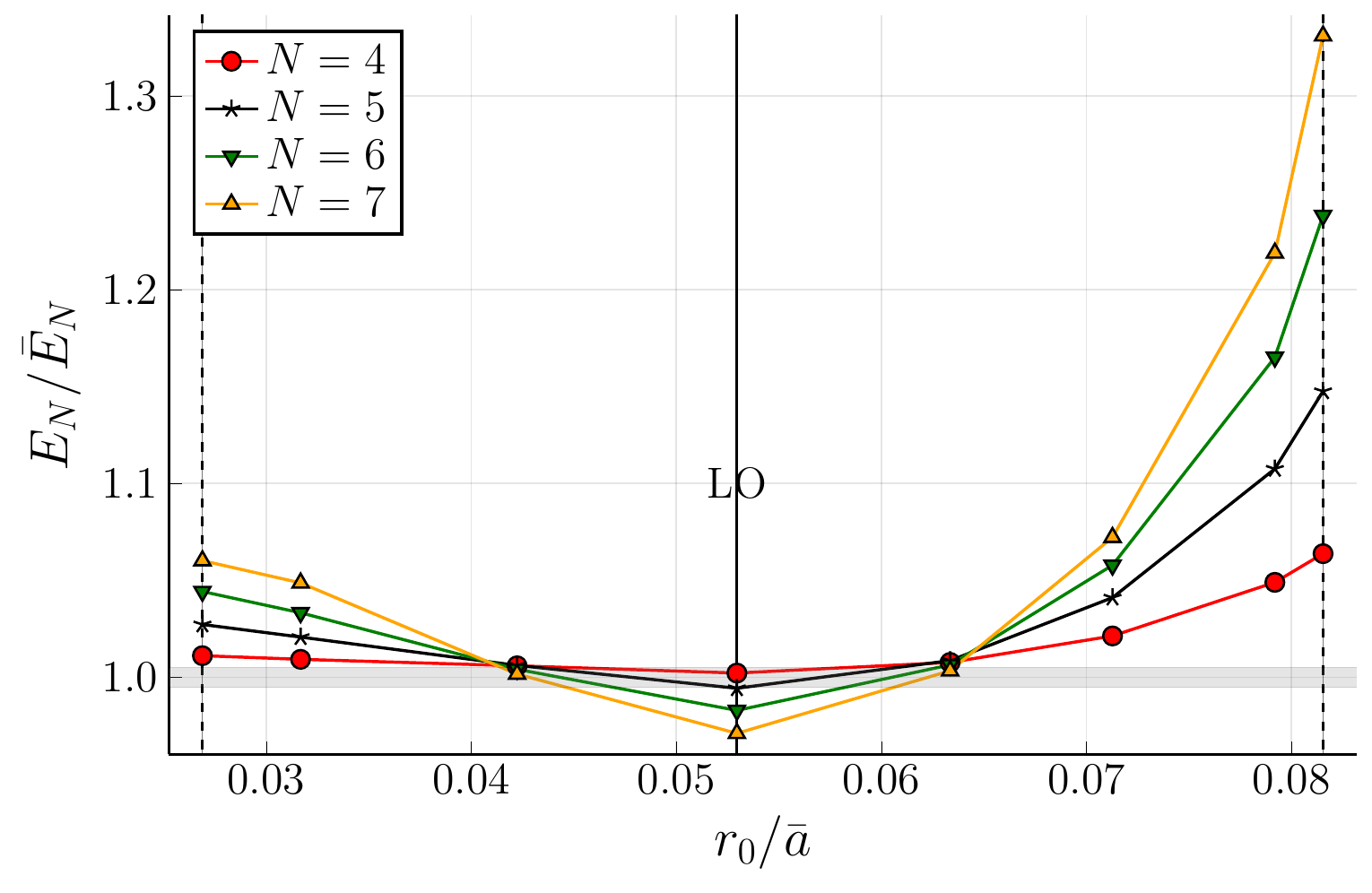}
  \end{center}
  \caption{The ratios $E_N/\bar E_N$ for $N=4,5,6,7$ as a
  function of $r_0/\bar a$ computed with the two-body force at NLO and
  the three body force at LO. The three-body force has been fixed 
  to have the best $E_4/\bar E_4$ ratio, which is always larger than 
  1. The vertical lines are the same as in Fig.~\ref{fig:runningNLO}. The 
  horizontal gray strip corresponds to the $\varepsilon^2=5\permil$ departure from LM2M2
  data. The point where $E_4/\bar E_4$ is closer to 1 is the 
  NLO-LO point given by Eq.~(\ref{eq:valuesLO}).}
  \label{fig:allNLOWith}
\end{figure}

\begin{figure}
  \begin{center} 
    \includegraphics[width=0.9\linewidth]{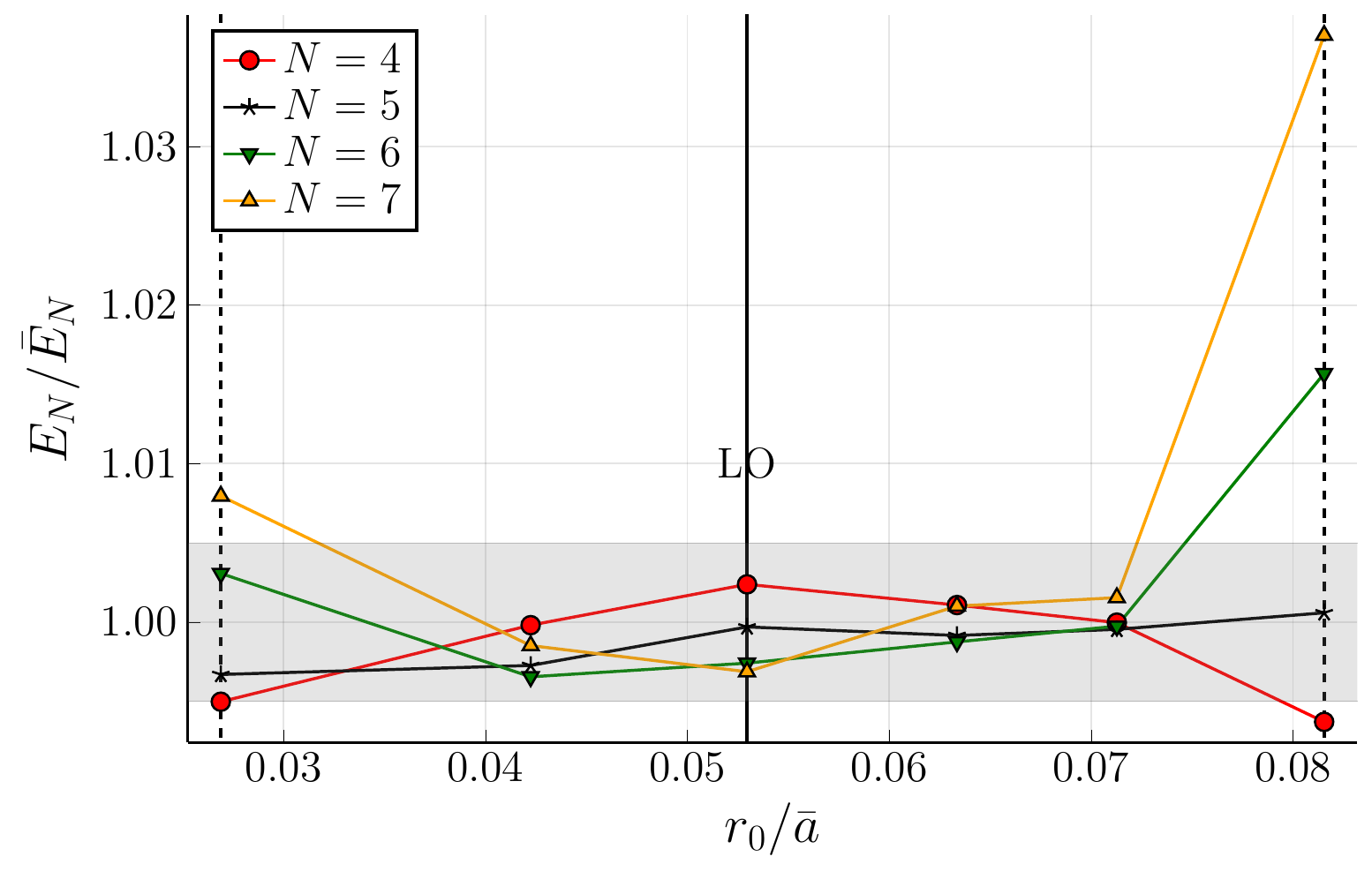}
  \end{center}
  \caption{The ratios $E_N/\bar E_N$ for $N=4,5,6,7$ as a
  function of $r_0/\bar a$ computed with the two-body and three-body
	 forces at NLO. The three-body force parameters have been fixed 
  to have $E_3/\bar E_3=1$ and the best $E_4/\bar E_4$ ratio. 
  The vertical lines are the same as in Fig.~\ref{fig:runningNLO}. The 
  horizontal gray strip corresponds to the $\varepsilon^2=5\permil$ departure from LM2M2
  data. The results stay inside the expected error band for values of
	$r_0/\bar a$ close to the LO vaue.}
  \label{fig:NLOfull}
\end{figure}

\begin{figure}
  \begin{center} 
    \includegraphics[width=0.7\linewidth]{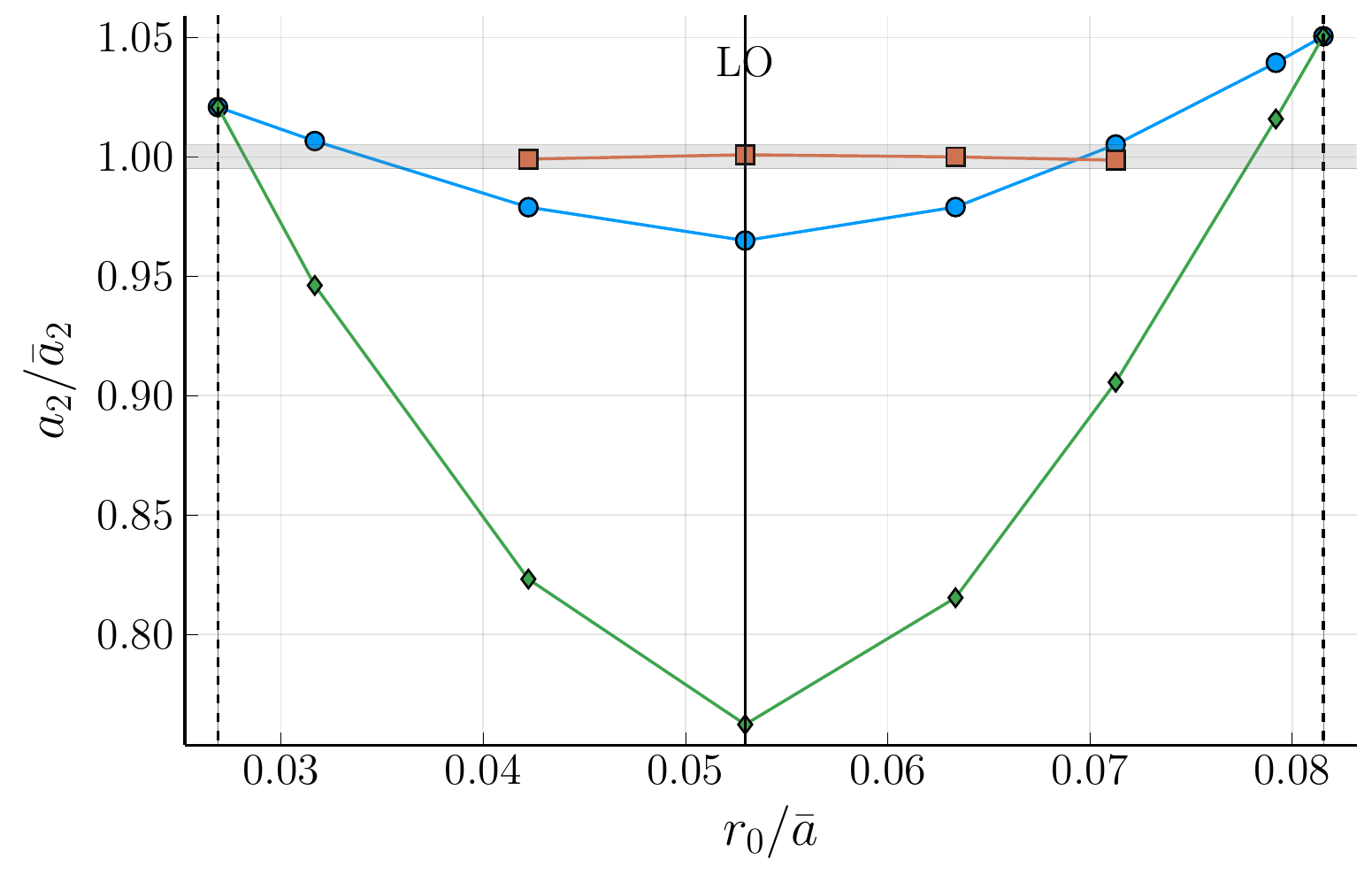}
  \end{center}
  \caption{The atom-dimer scattering length, $a_2$ (in units of the
  LM2M2 atom-dimer scattering length $\bar a_2 = 218.0~a_0$), 
  as a function of $r_0/\bar a$. The calculations have been done
  using the tewo-body NLO potential (green symbols), including the
  the LO three-body force (blue symbols) and the NLO three-body
	 force (red symbols).
  The vertical lines are the same as in
  Fig.~\ref{fig:runningNLO}, and the horizontal gray strip gives the
  $\varepsilon^2=5\permil$ deviation.}
  \label{fig:a2NLO}
\end{figure}
\section{Conclusions and Outlooks}~\label{sec:conclusions}

We have studied in detail how the leading order and the next to leading
order  interactions of a boson system can be built
by looking to a few data. In the LO case we have looked at the
two-body scattering length and the trimer energy whereas at NLO
we have considered the effective range and the tetramer binding energy
(or the atom-dimer scattering length). The ranges of the associated 
potentials (of a Gaussian shape) have been varied inside certain
regions. At LO we have shown that, if we want to maintain similar 
level of accuracy in the description of systems with increasing values of
$N$, the possible values of $r_0$ and $\rho_0$ are very few. In fact, as it is
evident from Fig.~\ref{fig:allLOWith}, there is only one possible pair
of values, $r_0$ and $\rho_0$, that respects this condition. Moreover
the value of $r_0$ is the one that allows a
simultaneously description of the two-body scattering length and of the
effective range. This is an important finding since this particular
value of $r_0$ gives the correct description of the two-pole structure
of the $S$-matrix showing the strict correlation that this structure
introduces in heavier systems. Moreover, associated to that particular $r_0$ value,
there is a particular value of the three-body potential range $\rho_0$ that takes 
into account the correct balance between attraction and repulsion along the energy curve $E_N$
as a function of $N$. That particular value of $\rho_0$ governs the
transition from universal to non-universal effects in which the
short-range characteristic of the interaction prevents the system to
collapse (see the related discussions in 
Refs.~\cite{kievsky:2017_Phys.Rev.A,kievsky:2018_Phys.Rev.Lett.,%
kievsky:2021_Phys.Rev.A,kievsky:2020_Phys.Rev.A}).

The LO description establishes the level of accuracy required in
the description of the binding energies. A consistent improvement is
expected once the next-to-leading order is considered. The first
finding in this analysis was the observation that the two-body NLO 
potential produces a maximum (in absolute value) in the description of the 
binding energies, $E_N$, with $N\ge 3$, located at the particular point in 
which the second LEC of the potential $V_1=0$. 
At this point the NLO potential consists in a
single Gaussian, and by definition, its range is the one needed to describe, in addition to the
scattering length, the effective range. As soon as $r_0$ is varied
from that value the trimer energy increases arriving to two values
at which the trimer energy is well described, $E_3=\bar E_3$.
Beyond these values the three-body force should be attractive, a
situation that without other intervening mechanisms would produce a collapse of the
system. So, in the present analysis we have studied the NLO force inside
those limits.

The next step in this study was the introduction of the three-body potential. To start with, in Fig.~\ref{fig:allNLOWith} we have analysed the
NLO two-body force in conjunction with the three-body force at LO.
This analysis shows a limited improvement of the description, mostly
around two points, at $r_0/\bar a \approx 0.042$ and $0.063$. At these
two points the description of the binding energies up to $N=7$ remains
inside the expected level of confidence. The final results are shown
in Fig.~\ref{fig:NLOfull} in which the complete NLO force has been
considered. Noticeably the complete segment of $r_0/\bar a$ values allows
for a description of the binding energies inside the error strip.
However, as mentioned before, at each point it was necessary to set the value of
$\rho_0$. So we have shown that there is a strict correlation between
the two-body range $r_0$ and the three-body range $\rho_0$. 
Similar conclusions are obtained analysing the results for the atom-dimer 
scattering length. This
fact has been observed at LO in fermionic
systems~\cite{kievsky:2018_Phys.Rev.Lett.,schiavilla:2021_Phys.Rev.C}.
Studies along the inclusion of the NLO term in such systems are at present under way.

\newpage
\appendix

\section{Special point where the two-body force at LO and NLO are the same}
We show why the binding energies calculated using the two-body force at NLO 
have their minimum value at the particular $r_0$ value in which $V_1=0$
(see Eq.(\ref{eq:nlo})). At that point the two-body potential is the
same at LO and NLO. To this end we first extend the Hellmann-Feynman
theorem 

\begin{equation}
 \frac{\partial E_\lambda}{\partial\lambda}
 = \langle\Psi_\lambda | \frac{\partial H_\lambda}{\partial\lambda} | \Psi_\lambda \rangle\,,
\end{equation}
\label{eq:HFT}
to be valid in the zero-energy case.

\subsection{Hellmann-Feynman theorem for the scattering length}
Here we demonstrate the validity of Hellmann-Feynman theorem in the
case in which the variation is done on the scattering length. 
We start from the Kohn variational
principle for the scattering length functional
\begin{equation}
[a_\lambda]=a_\lambda + 
\langle\psi_\lambda | H_\lambda | \psi_\lambda\rangle \,,
\end{equation}
where $[a_\lambda]$ is the second order estimate and $\psi_\lambda$ is
normalized so that $\psi_\lambda \rightarrow F+a_\lambda G$, with $F,G$ the
regular and irregular asymptotic solutions. Now the 
variation of the functional with respect to $\lambda$ is

\begin{equation}
 \frac{\partial [a_\lambda]}{\partial\lambda} =
 \frac{\partial a_\lambda}{\partial\lambda} +
  \langle\frac{\partial \psi_\lambda}{\partial\lambda} | H_\lambda| \psi_\lambda \rangle +
  \langle\psi_\lambda | \frac{\partial H_\lambda}{\partial\lambda} | \psi_\lambda \rangle+
  \langle\psi_\lambda |H_\lambda| \frac{\partial \psi_\lambda}{\partial\lambda}\rangle \,.
\end{equation}
From the normalization condition we can observe that
\begin{equation}
  \langle\frac{\partial \psi_\lambda}{\partial\lambda} | H_\lambda| \psi_\lambda \rangle -
  \langle\psi_\lambda |H_\lambda| \frac{\partial \psi_\lambda}{\partial\lambda}\rangle =
  \frac{\partial a_\lambda}{\partial\lambda}
\end{equation}
and therefore
\begin{equation}
 \frac{\partial [a_\lambda]}{\partial\lambda} =
  2\langle\frac{\partial \psi_\lambda}{\partial\lambda} | H_\lambda| \psi_\lambda \rangle +
  \langle\psi_\lambda | \frac{\partial H_\lambda}{\partial\lambda} | \psi_\lambda \rangle=
  \langle\psi_\lambda | \frac{\partial H_\lambda}{\partial\lambda} | \psi_\lambda \rangle
  \label{eq:HF_a0}
\end{equation}
where we have used that $H_\lambda| \psi_\lambda \rangle=0$.
Accordingly, the extension of the Hellmann-Feynman theorem to the case of
a zero-energy state results to be:
\begin{equation}
 \frac{\partial [a_\lambda]}{\partial\lambda} =
 \langle\psi_\lambda | \frac{\partial H_\lambda}{\partial\lambda} | 
	\psi_\lambda \rangle
  \label{eq:HF_a}
\end{equation}

\subsection{Minimum of $V_0$ for $V_1=0$.}
It is interesting to analyse why the minimum value of $E_N$ corresponds to the
case in which the LO and NLO two-body potentials are the same potential. 
In particular we observe 
that at that point $V_0$ has its minimum
value. We start from the two-body NLO hamiltonian
\begin{equation}
 H^{\text{NLO}}_{r_0}= T+ V^{\text{NLO}}\,,
\end{equation}
where we have made explicit the $r_0$ dependence. Now we recall the Hellmann-Feynman 
theorem 
\begin{equation}
 \frac{\partial E_\lambda}{\partial\lambda}
 = \langle\Psi_\lambda | \frac{\partial H_\lambda}{\partial\lambda} | \Psi_\lambda \rangle\,,
\end{equation}
for a general dependence of the hamiltonian on the parameter $\lambda$ and the
corresponding wave function $\Psi_\lambda$. Its
extension to the case of the scattering length is
\begin{equation}
 \frac{\partial a_\lambda}{\partial\lambda}
 = \langle\psi_\lambda | \frac{\partial H_\lambda}{\partial\lambda} | \psi_\lambda \rangle\,,
\end{equation}
where now $\psi_\lambda$ is the zero-energy wave function (a demonstration of
the theorem for scattering states is given above). In our case 
$\lambda\equiv r_0$ and due to the constrains, $E_2$ and $a$
are maintained constants as $r_0$ is varied, the variations of $E_2$ and $a$
with respect to $r_0$ are zero. Explicitly
\begin{equation}
 \begin{aligned}
 \frac{\partial E_2}{\partial r_0} &=0
 = \langle\Psi_{r_0} | \frac{\partial H^{\text{NLO}}_{r_0}}{\partial r_0} | \Psi_{r_0} \rangle 
 = \langle\Psi_{r_0} | \frac{\partial V^{\text{NLO}}}{\partial r_0} | \Psi_{r_0} \rangle  \\
 &= \langle\Psi_{r_0} | 
\left( \frac{\partial V_0}{\partial r_0} + 2 V_0 \frac{r^2}{r_0^3}\right ) e^{-(r^2/r_0^2)}+
\left( \frac{\partial V_1}{\partial r_0} \frac{r^2}{r_0^2}- 2 V_1 \frac{r^2}{r_0^3}
+ 2 V_1 \frac{r^4}{r_0^5}\right ) e^{-(r^2/r_0^2)}
| \Psi_{r_0} \rangle \,,
 \end{aligned} 
\end{equation}
where $\Psi_{r_0}$ is the bound state wave function.
We are interested in the point in which $V_1=0$, therefore the equation is
\begin{equation}
 \frac{\partial V_0}{\partial r_0} \langle\Psi_{r_0} |e^{-(r^2/r_0^2)}|\Psi_{r_0}\rangle +
\left( \frac{2 V_0}{r_0} + \frac{\partial V_1}{\partial r_0} \right) 
 \langle\Psi_{r_0} |\frac{r^2}{r_0^2} e^{-(r^2/r_0^2)}+ | \Psi_{r_0} \rangle=0\,,
\end{equation}
and a similar equation for the zero-energy wave function, $\psi_{r_0}$
\begin{equation}
 \frac{\partial V_0}{\partial r_0} \langle\psi_{r_0} |e^{-(r^2/r_0^2)}|\psi_{r_0}\rangle +
\left( \frac{2 V_0}{r_0} + \frac{\partial V_1}{\partial r_0} \right) 
 \langle\psi_{r_0} |\frac{r^2}{r_0^2} e^{-(r^2/r_0^2)}+ | \psi_{r_0} \rangle=0\,.
\end{equation}

Since $\Psi_{r_0}$ and $\psi_{r_0}$ are orthogonal, the gaussian matrix elements cannot be
linear dependent and therefore the only possibility is that each coefficient is
zero. In particular the following condition is verified
\begin{equation}
 \frac{\partial V_0}{\partial r_0}=0 \,,
\end{equation}
together with $2 V_0/r_0 + \partial V_1/\partial r_0=0$. These
conditions indicate that when the variation is performed on $E_N$
it results
\begin{equation}
\frac{\partial E_N}{\partial r_0}=0
\end{equation}
at the $V_1=0$ point. This behavior can be observed in Fig.\ref{fig:NLOWo}
for $N=3,4,5,6$.

\begin{acknowledgments}
  This research was supported in part by the National Science Foundation under
  Grant No. NSF PHY-1748958.
\end{acknowledgments}

\clearpage
\bibliography{biblio}   

\end{document}